\documentclass[journal]{IEEEtran}
\usepackage{amsmath,amsfonts}
\usepackage{algorithmic}
\usepackage{algorithm}
\usepackage{array}
\usepackage[caption=false,font=normalsize,labelfont=sf,textfont=sf]{subfig}
\usepackage{textcomp}
\usepackage{stfloats}
\usepackage{url}
\usepackage{verbatim}
\usepackage{graphicx}
\usepackage{cite}
\usepackage{booktabs}
\usepackage{multirow}
\usepackage{siunitx}
\usepackage{hyperref}
\usepackage{xcolor}
\begin{document}

\title{Perceptual Dimensions of Physical Properties of Handheld Objects Induced by Impedance Control of Human Motion\thanks{This paper has been published in IEEE Access: \url{https://doi.org/10.1109/ACCESS.2025.3553042}}}

\author{
Takeru Hashimoto,
Shigeo Yoshida,
Takuji Narumi

\thanks{Manuscript received April 19, 2021; revised August 16, 2021.}
}

\markboth{Journal of \LaTeX\ Class Files,~Vol.~14, No.~8, August~2021}%
{Shell \MakeLowercase{\textit{et al.}}: A Sample Article Using IEEEtran.cls for IEEE Journals}


\maketitle

\noindent
\fcolorbox{red}{white}{
\begin{minipage}{0.9\columnwidth}
\textcolor{red}{
\textbf{Note:} The final version of this paper has been published in IEEE Access:\\
\url{https://doi.org/10.1109/ACCESS.2025.3553042}
}
\end{minipage}
}

\begin{abstract}
Haptics in virtual reality is a new frontier beyond audiovisual experiences. 
Researchers are designing handheld VR controllers to simulate haptic sensations.
However, such research often used a single measure to assess user perception, potentially overlooking other sensory experiences.

Therefore, this study developed a haptic display and evaluates not \textit{how much} but \textit{how} humans feel a physical property when haptic stimuli are changed.
We conducted interviews to investigate how people feel when a haptic device changes motion impedance.
We used thematic analysis to abstract the results of the interviews and gain an understanding of how humans attribute force feedback to a phenomenon.
We also generated a vocabulary from the themes obtained from the interviews and asked users to evaluate force feedback using the semantic difference method.
A factor analysis was used to investigate how changing the basic elements of motion, such as inertia, viscosity, and stiffness of the motion system, affects haptic perception.
As a result, we obtained four critical factors: size, viscosity, weight, and flexibility factor, and clarified the correspondence between these factors and the change of impedance.
\end{abstract}

\noindent
\fcolorbox{red}{white}{
\begin{minipage}{0.9\columnwidth}
\textcolor{red}{
\textbf{Note:} The final version of this paper has been published in IEEE Access:\\
\url{https://doi.org/10.1109/ACCESS.2025.3553042}
}
\end{minipage}
}

\section{Introduction}
\IEEEPARstart{H}{aptics} is increasingly recognized as a crucial dimension in the realm of virtual reality, alongside the well-established audiovisual aspects.
In recent years, researchers have been exploring various handheld VR controllers to express haptic sensations to users within virtual environments.
Among these, active haptic interfaces stand out as promising methods capable of dynamically shaping the user's haptic experiences by actively modulating the force feedback they provide.

In active haptic interfaces, the key variable is not a static physical property but a time-varying relationship between the force input by the user and the corresponding reaction force from the device. 
However, it's essential to recognize that the fidelity of forces produced by such devices, while accurate, may not replicate the richness of information present in the real world. 
This discrepancy arises due to the mechanical nature of force generation and the intricate relationship between physical phenomena and human perception.

Assessing human perception in the context of haptic devices often relies on psychophysical experiments that allow researchers to convey how well a haptic device can convey specific physical properties or sensations.
However, the limitation of such experiments lies in their inherently one-dimensional nature, determined by the experimenter's predefined criteria.
For instance, asking participants to evaluate force feedback as an object's ``weight'' confines the assessment to a single aspect of haptic perception, potentially neglecting other sensations like ``stickiness'' or ``stiffness.'' 
This limitation reflects an inherent top-down imposition of ``reality'' onto the user's sensory experience.

Therefore, as a preliminary step to such a psychophysical \textit{quantitative} assessment, we need to \textit{qualitatively} investigate the possible perceptual phenomena that can be represented by the haptic device. 
This investigation will not only allow us to explore the representational possibilities of the various haptic devices that have been created, but also to determine how independently each representation can be controlled.
Regarding cutaneous sensation, some studies have investigated the correspondence between the physical properties of surfaces and human perception.
They have extracted independent and significant factors related to tactile perception and construct perceptual dimensions of tactile surface texture~\cite{Shirado2005-qk,Bergmann_Tiest2006-qg}.
The correspondence in proprioceptive sensation should also be investigated for the future design of haptic devices.

In particular, our study focuses on the relationship between the impedance control parameters of a force feedback device and the perception of physical properties associated with a handheld object. 
By using a single haptic device using gyro moment, we aim to demonstrate how active force feedback affects human haptic perception and to clarify the relationship between the physical and perceptual domains.

By conducting this study, we hope to contribute to the broader field of haptic technology by gaining insights into the intricacies of human haptic perception and advancing the design and utilization of haptic interfaces for a more immersive and sensory-rich virtual reality experience.

The contribution of this paper as follows:
\begin{enumerate}
    \item Development and evaluation of the ungroundedd one-DoF torque feedback device and impedance control system
    \item Qualitative exploration of the sense of physical properties that impedance control can present based on thematic analysis
    \item Construction of perceptual dimensions of physical properties of handheld objects induced by impedance control
\end{enumerate}

\section{Related Work}\label{sec:review}
This article draws inspiration from various studies in haptic feedback for representing object properties, evaluating human perception of material, and ecological psychology.
\subsection{Dynamic Touch}
In ecological psychology, research has extensively explored how humans perceive the properties of handheld objects through haptic perception.
These studies have demonstrated the critical role of haptic sensation in perceiving various properties of handheld objects, including length, width~\cite{Turvey1998}, and three-dimensional shape~\cite{burton1990can}.
As the term ``dynamic touch "~\cite {Turvey1996-ml} suggests, this wealth of information is acquired through dynamic manipulation of the object rather than static handling. 
Humans infer a plausible object shape from the forces applied to tendons and muscles in the arm and the skin deformation on the palm, achieved through various interactions with the object.

\subsection{Haptic Exploration of Objects}

Haptic exploration is a pattern of behavior that individuals intentionally engage in to characterize surfaces and objects, a spontaneous pattern that optimizes the uptake of information.
Lederman et al. described a series of specialized search patterns called exploratory procedures (EPs)~\cite{lederman1987hand}.
Among EPs, \textit{unsupported holding} is particularly effective in acquiring information on the properties of the handheld object.
While their paper discussed only weight as an associated property, later experiments showed that humans could perceive various object properties by \textit{unsupported holding}, in their words, \textit{dynamic touch}~\cite{Turvey1995-kn}.
Turvey et al. focused on information obtained from grasping and swinging. 
The properties of an object obtained by swinging include length, weight, width, shape of the object tip, and direction of the hand relative to an object. 
This information is derived from the sensitivity of body tissues to the values of rotational torque and rotational dynamics.

\subsection{Relationship between Physical Properties and Perception}
Many studies have explored the connection between materials and perceptual information using cutaneous touch~\cite{Hollins1993-eu, Bergmann_Tiest2006-qg, Shirado2005-qk}.
For instance, in a study using a multi-dimensional scaling method on 124 material samples, researchers identified that humans rely on four primary factors to assess material quality~\cite{Bergmann_Tiest2006-qg}.
The extent to which each physical quantity affects the sensation scale has also been investigated through multivariate analysis~\cite{Shirado2005-qk}.
In contrast, limited research has been conducted to evaluate the relationship between the properties of handheld objects and the perceptual information acquired through dynamic touch. 
As mentioned in the introduction, one-dimensional measurements have shown that the moment of inertia and rotational dynamics correspond to the perception of length and width~\cite{Turvey1998,burton1990can, Park2023-kr}. 
However, they do not guarantee that other factors are unaffected.

In the context of dynamic touch, it is crucial to examine the connection between physical properties and perceptual information.
This correspondence can be used as an indicator to determine which physical properties should be manipulated to achieve the desired perceived information.
In addition, only by understanding the correspondence between physical properties and perceptions will it be possible to conduct psychophysical experiments that will yield as much information as possible.

\subsection{Passive Haptic Interfaces}
To represent various physical properties of a handheld object, researchers have developed a variety of haptic interfaces. 
In this section, we will focus on the category known as ``passive haptics.''
This method represents the physical properties of a handheld object by changing the physical properties of the interface itself.
There is research on haptic proxies, which create virtual objects by combining small artifacts or adjusting the weight position~\cite{fujinawa2017computational, Zhu2019-aa, Chen2021-rg, 9284771}.
However, these require reassembling the proxies each time the virtual object changes.
On the other hand, the field also explores \textit{dynamic} passive haptic interfaces. 
These interfaces use actuators not for dynamic force output, but to change the actual static physical properties of the device.
These devices incorporate multiple actuators within a single unit to modify the center of gravity~\cite{Swindells_2003, Zenner_2017,shigeyama2019transcalibur}, inertia~\cite{Zenner_2017,shigeyama2019transcalibur}, shape~\cite{Gonzalez_Avila2021-ll}, drag~\cite{zenner2019drag,10.1145/3305367.3327991}, stiffness~\cite{Ryu2020-el, Tsai2020-ea, Ryu2021-xi}, and other physical properties of the interface itself.
Unlike haptic proxies, these devices do not require reassembly.
However, these devices are often specialized for rendering a specific physical property, and the number of actuators increases when we want to change multiple properties simultaneously.

\subsection{Active Haptic Interfaces}
This study focuses on active haptic interfaces, devices that employ actuators to generate force output directly.
Active haptic interfaces are categorized based on where the reactive force to the generated force is received: environment-grounded, body-grounded, and ungrounded. 
We focus on the ungrounded category, which is most closely related to our research.
Ungrounded haptic interfaces vary in the methods they employ to achieve force feedback.
Some devices utilize the force of propeller rotation to push air~\cite{heo2018thor,10.1145/3332165.3347926, Sasaki:2018:LMH:3214907.3214913, Ke2023-qk} or the reaction force of compressed air release~\cite{Wang2021-sd, Tsai2022-sz}. 
However, these methods often suffer from disadvantages such as force output delay and imprecise force output.
On the other hand, devices that utilize the gyro effect to output force have also been studied~\cite{Winfree2009-rv, Walker_2018, nakamura2005innovative, 1191223, Hashimoto2022-af, Hashimoto2023-ah}. 
Those methods have a slight delay and are characterized by rotational torque output rather than translational force.

Instead of generating actual force, researchers have found that stimulating muscles with electrical stimulation can be used to generate force feedback~\cite{Lopes2017, Lopes2015}.
While EMS devices are small and lightweight, they are challenging to control force precisely.
Researchers have demonstrated the use of rapidly moving weights to simulate force output~\cite{Shimizu2021-df} or the use of asynchronous periodic oscillation to create the sense of pulling or pushing due to the nonlinear nature of our haptic perception~\cite{4600289}.

Consequently, a wide array of haptic devices has been developed, each with advantages and disadvantages. 
While this paper employs a haptic device utilizing gyro-moments, the impedance control method employed here can be adapted to haptic devices using alternative methods. 
Thus, the insights into the haptic dimension gained in this study have broader applicability across various systems.

\section{Implementation of A Haptic Device}
We aim to consider the design requirements for a device necessary to investigate the dimensions of property perception of the handheld object.
The design requirements are as follows:
\begin{enumerate}
    \item Having a mechanism capable of ungrounded force output.
    \item Capable of low-latency force feedback and impedance control synchronized with human movements.
\end{enumerate}

The necessity of ungrounded characteristics, as stated in 1), is crucial.
In scenarios involving shaking or manipulating an object, the ability of a force feedback device to output force without grounding is highly significant. 
In contrast to traditionally grounded manipulators, ungrounded devices possess higher portability, allowing free movement without restricting the workspace. 
Grounded manipulators tend to have lower portability, necessitating expensive torque sensors for unrestricted human movement. 
Hence, ungrounded is immensely beneficial in the property perception of a wielded object.
The rationale for the essentiality of low latency, as mentioned in 2), is expounded. 
The impedance control utilized in this study demands immediate force feedback based on the angular acceleration and velocity of the device resulting from human body movements. 
In a control system with significant delays in force output, users would find it challenging to infer the relationship between their movements and the force feedback. 
Therefore, low-latency characteristics play a vital role in unraveling the dimensions of property perception of a wielded object.

To meet the aforementioned requirements, We employ gyroscopic moment as the force feedback method. 
As discussed in section~\ref{sec:review}, gyroscopic moment facilitates ungrounded and low-latency force output. 
The rotational torque output generated by this device aligns well with human perceptual characteristics, as humans perceive properties of a wielded object by inducing rotational movements.

\begin{figure}
    \centering
    \includegraphics[width=\linewidth]{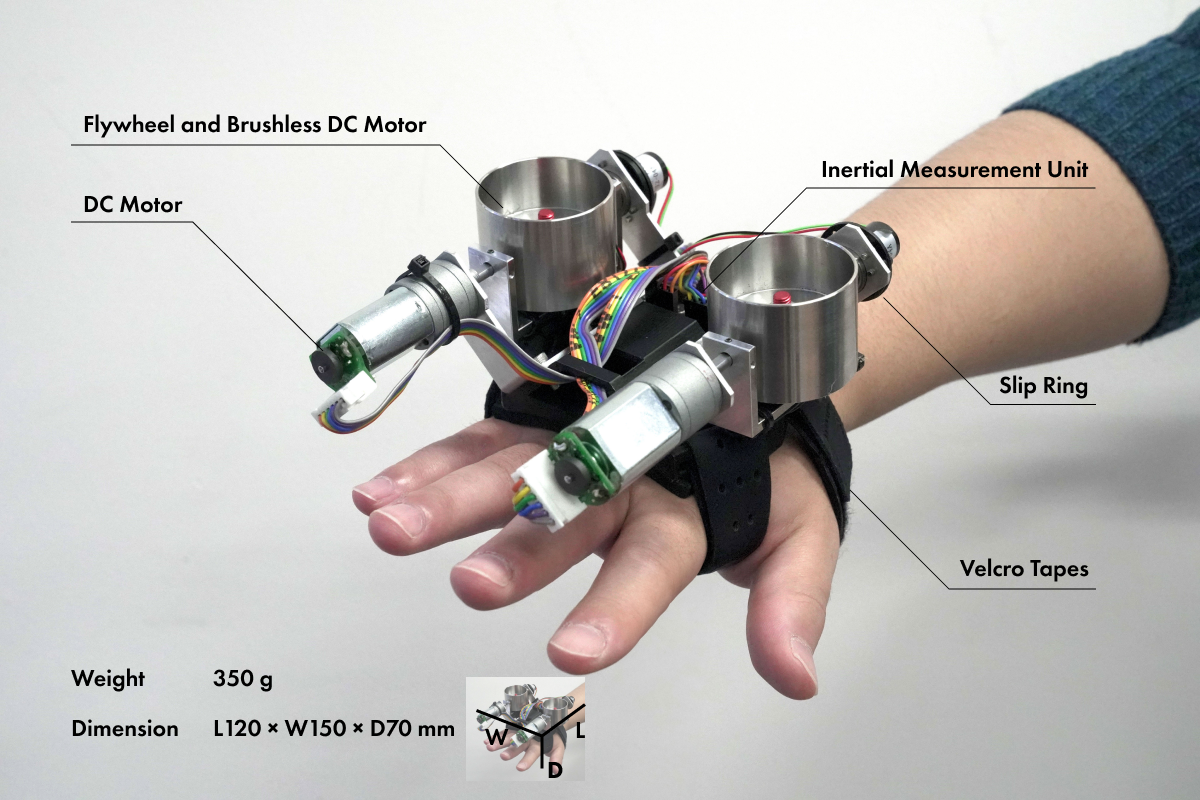}
    \caption{The device consists of two CMGs attached to a base component, and it can be worn on the body using a pair of intersecting Velcro tapes.}
    \label{fig:hardware}
\end{figure}

\begin{figure}
    \centering
    \includegraphics[width=\linewidth]{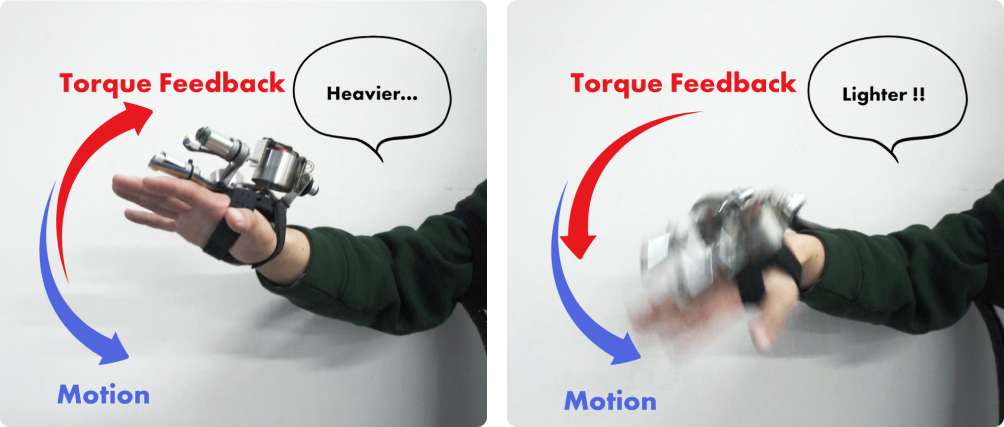}
    \caption{The direction of the user's motion (blue arrow) and the direction of the device's torque output (red arrow). Both are in the pitch direction. The positive, negative, and magnitude of the torque changes the physical properties perceived by the user.}
    \label{fig:pitch}
\end{figure}

We built a wearable haptic device that consists of mechanisms to output torque by tilting the spinning flywheel called control moment gyroscopes (CMG) based on a previous study~\cite{Hashimoto2023-ah}.
Figure~\ref{fig:hardware} shows mechanism models and specifications of the device.
This device can output a torque with one degree of freedom.
The direction of the torque that can be output varies depending on the initial orientation of the flywheel.
In this paper, the torque was output in the pitch direction  (see Figure \ref{fig:pitch}), which would be most used to perceive the physical properties of the grasped object.
Weight is an important form factor in wearable devices. 
After repeated prototyping, the final weight was set to \SI{350}{g}.


\subsection{Hardware Implementation}
\subsubsection{Flywheel} 
The flywheel is fabricated from stainless steel to ensure high density, and it possesses a diameter of \SI{40}{mm}, a mass of \SI{30}{g}, and a moment of inertia around the rotational axis $I$ of \SI{1.27 e-5}{kgm^2}. 
A brushless DC motor is employed to spin the flywheel , which consistently rotates around \SI{8000}{rpm} (defined as $\Omega$).

\subsubsection{Gimbal Mechanism} 
We utilized a Pololu 63:1 geared DC motor and its encoder to drive the rotation of the gimbal mechanism. 
A slip ring was installed on the motor's opposite side to prevent cable entanglement.

\subsubsection{Structure}
The flywheel and gimbal bases are constructed from aluminum for strength and lightness, while the part worn on the hand is 3D-printed using PLA. 
The device can be securely attached to a user's hand by two intersecting Velcro tapes.

\subsubsection{IMU} The device is equipped with an Inertial Measurement Unit (IMU) to monitor angular velocity and orientation, and angular acceleration is calculated by differentiating the measured angular velocity from the IMU.

\subsection{Torque Calculation}
\begin{figure}
    \centering
    \includegraphics[width=\linewidth]{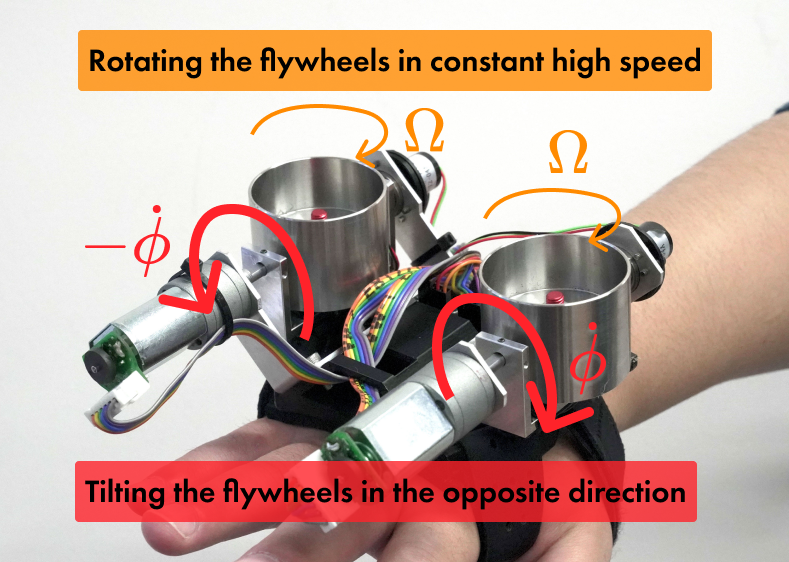}
    \caption{Direction of rotation of the flywheels and each gimbal. 
    The flywheels rotate at a constant speed $\Omega$ and the two gimbals each move at speed $\dot\phi$ in opposite directions to tilt the flywheels.}
    \label{fig:acutuation}
\end{figure}

The torque output $\tau$ in CMG systems is determined by the angle $\phi$ and angular velocity $\dot\phi$ of the gimbals.
Our device uses a scissored paired CMG configuration, where two CMGs are positioned in parallel, with each gimbal rotating in opposite phases as figure \ref{fig:acutuation}.
That is, the angle of each gimbal is represented as $(\phi_1,\phi_2) = (\phi,-\phi)$, and the torque output $\tau$ is as follows:
\begin{equation}
    \tau = (-2\sin\phi,0, 0)^T \dot\phi
\end{equation}
The velocity of the gimbal $\dot\phi$ to obtain the desired torque $\tau$ is calculated from the inverse problem of this equation.

\subsection{Impedance Control}
The concept of impedance control is to represent an object's static properties as force information.
This concept was proposed so that multi-axis robot arms can output an appropriate force when they come into contact with humans~\cite{hogan1985impedance}.
In impedance control, the force output system adjusts the force output according to acceleration, velocity, and displacement so that the three desired values of \textit{inertia}, \textit{damping}, and \textit{stiffness} in the equation of motion are achieved.

Prior studies have considered inertia and damping coefficients as impedance control variables when they represent the properties of handheld objects~\cite {Hashimoto2022-af}.
However, there are several examples of varying stiffness in prior cases~\cite{Ryu2020-el, Tsai2020-ea, Ryu2021-xi}.
This study addresses all of the aforementioned impedance control parameters: mass, damping, and stiffness.
For rendering mass and damping, we refer to the previous study~\cite {Hashimoto2022-af}.
For rendering stiffness, we refer the model of two masses adopted in the Elastick~\cite{Ryu2020-el}.
Specifically, we define orientations for the two masses, base and tip, and implement the restorative force generated by the difference in orientations.
This allows us to capture the time lag between the motion at the base and the tip. 
The algorithm for calculating the torque $T_{e}$ due to the restorative force is presented in Algorithm~\ref{alg:alg1}.
In this algorithm, $k_r$ represents the constant of the restorative force applied to the rotational displacement, while $c_r$ is the damping coefficient applied to the difference in rotational velocity. $I$ represents the overall moment of inertia. Additionally, $\theta_1$ and $\theta_2$ denote the orientations of each mass.

\begin{figure}
    \centering
    \includegraphics[width=0.7\linewidth]{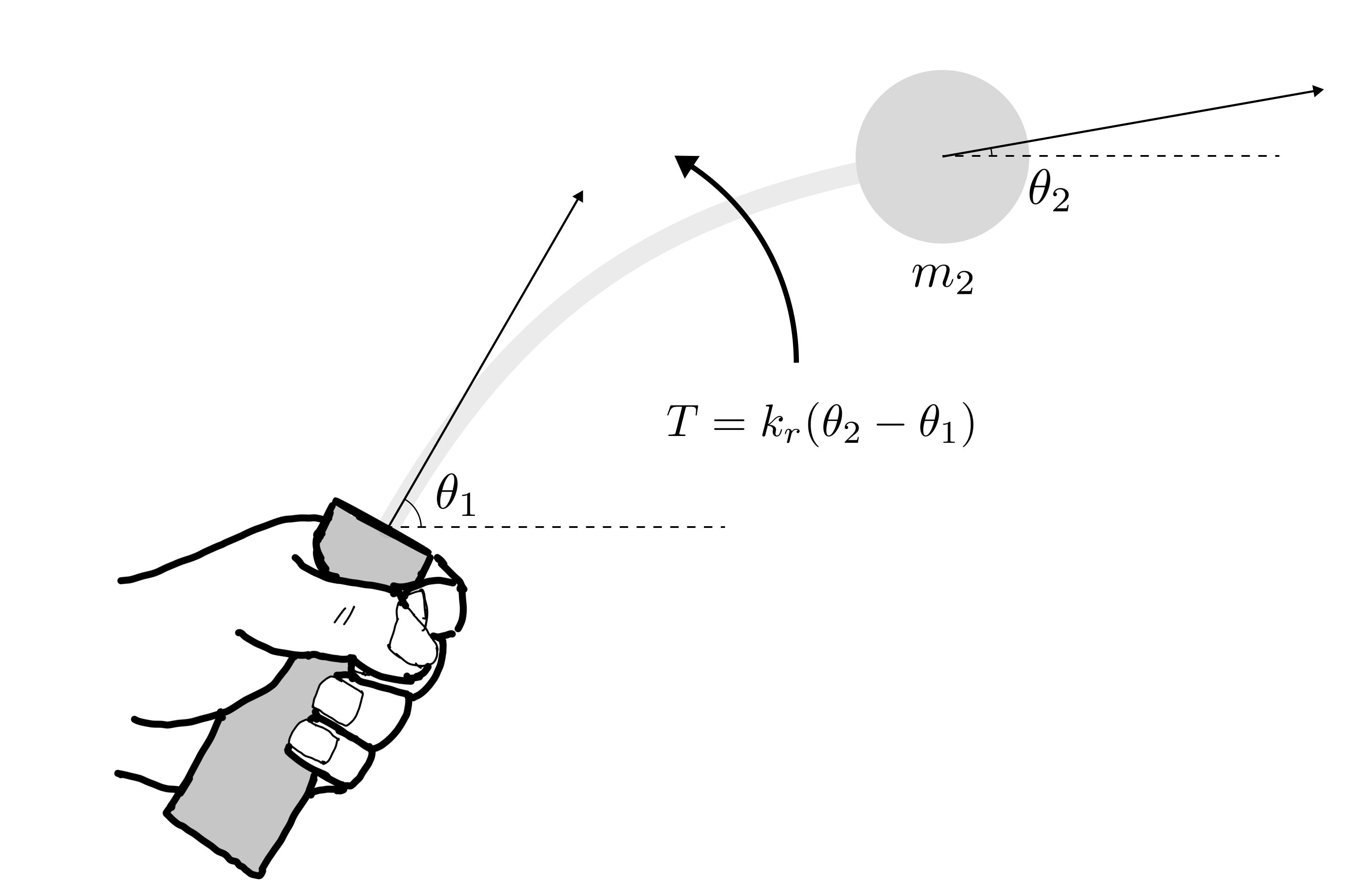}
    \caption{Bending restorative force model.}
    \label{fig:elastic model}
\end{figure}

\begin{algorithm}
\caption{Simulation of torque from elastic object.}\label{alg:alg1}
\begin{algorithmic}

    \STATE ${\theta_{1,2},\dot\theta_{1,2},\ddot\theta_{1,2}} =  0 \in \mathbf R^{3\times 1}$
    \STATE $k_r, c_r, I = \mathbf{const}.$
    \WHILE{simulation running}
    \STATE \#STEP 1: updating the state of the grip
    \STATE ${\ddot{\theta_1}}[t] = \ddot\theta_{imu}$ 
    \STATE ${\dot\theta_1}[t] += {\ddot{\theta_1}}[t]\Delta t~~~ {\theta_1}[t] += {\dot\theta_1}[t] \Delta t$

    \STATE \#STEP 2: calculating bending of the virtual rod
    \STATE $\Delta \theta = \theta_2 - \theta_1 ~~~ \Delta \dot\theta = \dot\theta_2 - \dot\theta_1$ 

    \STATE \#STEP 3: calculating bending torque
    \STATE $T_{e} = - k_r \Delta \theta - c_r \Delta \dot\theta$ 

    \STATE \#STEP 4: updating the state of the tip object
    \STATE ${\ddot{\theta_2}}[t] = I^{-1} T_{e}$
    \STATE ${\dot\theta_2}[t] += {\ddot{\theta_2}}[t]\Delta t~~~ {\theta_2}[t] += {\dot\theta_2}[t] \Delta t$
    
    \ENDWHILE
\end{algorithmic}
\label{alg1}
\end{algorithm}

Based on the above, the impedance control of the handheld object in this paper can be expressed by the equation~\ref{eq:rot} and \ref{eq:imp_rotate}.
Here, $\Delta I$ represents the difference in inertia, $\Delta D$ represents the difference in damping coefficient, and $\omega$ is the angular velocity of the device. 
It should be noted that the inherited damping coefficient $D$ and restorative force $k_r$ in the motion are assumed to be very close to zero.

\begin{equation}
    I\dot\omega = \tau_h + \tau_{gen}
    \label{eq:rot}
\end{equation}
\begin{equation}
    \tau_{gen} = -\Delta I\dot\omega  -\Delta D\omega  - T_{e}(k_r,c_r)
    \label{eq:imp_rotate}
\end{equation}

In summary, while previous studies have primarily focused on inertia and damping coefficients, this study explores the incorporation of elastic forces into impedance control, providing a more comprehensive framework for understanding object dynamics.

\section{Measurement of Generated Torque}\label{sec:measurement}
We conducted the torque measurement to verify whether the desired torque calculated by Equation ~(\ref{eq:imp_rotate}) could be output through the gimbal drive, based on the previous study~\cite{Hashimoto2022-af}.

\subsection{Measurement Setup}
\begin{figure}
    \centering
    \includegraphics[width = \linewidth]{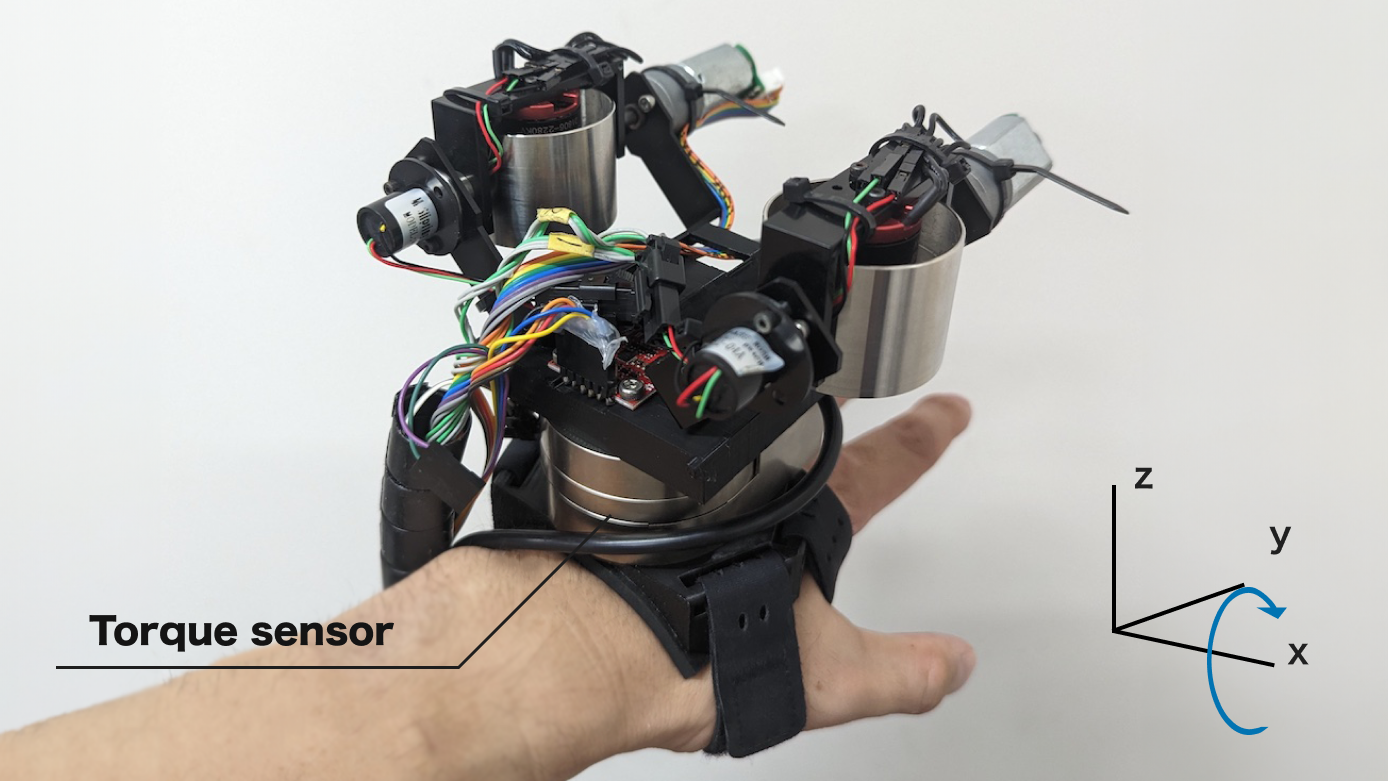}
    \caption{haptic device with the torque sensor was attached to the participant's hand. Participants performed a swing movement around the x-axis.}
    \label{fig:torque_fixed_body}
\end{figure}

Because the proposed system outputs torque in response to human motion, we measured the torque generated when the device was swung by a human. 
Figure~\ref{fig:torque_fixed_body} shows the device attached to the body with the torque sensor.
A collaborator performed the swing movement in the direction of wrist flexion and extension, x-axis in Figure~\ref{fig:torque_fixed_body}.
The variables for impedance under each condition were set as shown in Table~\ref{tab:parateters}.
Regarding the exclusion of the ``elasticity decrease'' condition, the low elasticity condition ensures that the object's tip remains in sync with the base's movement, eliminating noticeable lag. Thus, the effects of an ``elasticity decrease'' are inherently accounted for within the ``increased inertia'' condition. Consequently, considering ``elasticity decrease'' as a separate condition was unnecessary.

As defined in algorithm~\ref{alg:alg1}, there are two variables that influence the stiffness of the object. 
One is the spring coefficient restoring force proportional to the displacement of the virtual mass, and the other is the damping force proportional to the velocity of the virtual mass. 
The oscillation of an object is determined by these two variables, where a higher spring coefficient leads to higher-frequency vibrations and a higher damping coefficient results in faster convergence of vibrations.
Of course, these combinations can represent a variety of vibrations, but this paper focuses on one of them.

\begin{table}[htb]
\centering
\caption{Impedance parameters in each condition.}
\begin{tabular}{rrrrr}
\toprule
Condition & $\Delta I$ & $\Delta D$ & $(k_r, c_r)$ \\
\midrule
increased inertia & 0.002 & 0 & (0, 0)  \\
decreased inertia & -0.002 & 0 & (0, 0)  \\
damping increase & 0 & 0.02 & (0, 0) \\
damping decrease & 0 & -0.02 & (0, 0)  \\
elasticity increase & 0 & 0 & (0.2, 0.001)  \\
\bottomrule
\end{tabular}
\label{tab:parateters}
\end{table}

\subsection{Measurement Result}
\begin{figure}[]
    \centering
    \includegraphics[width=\linewidth]{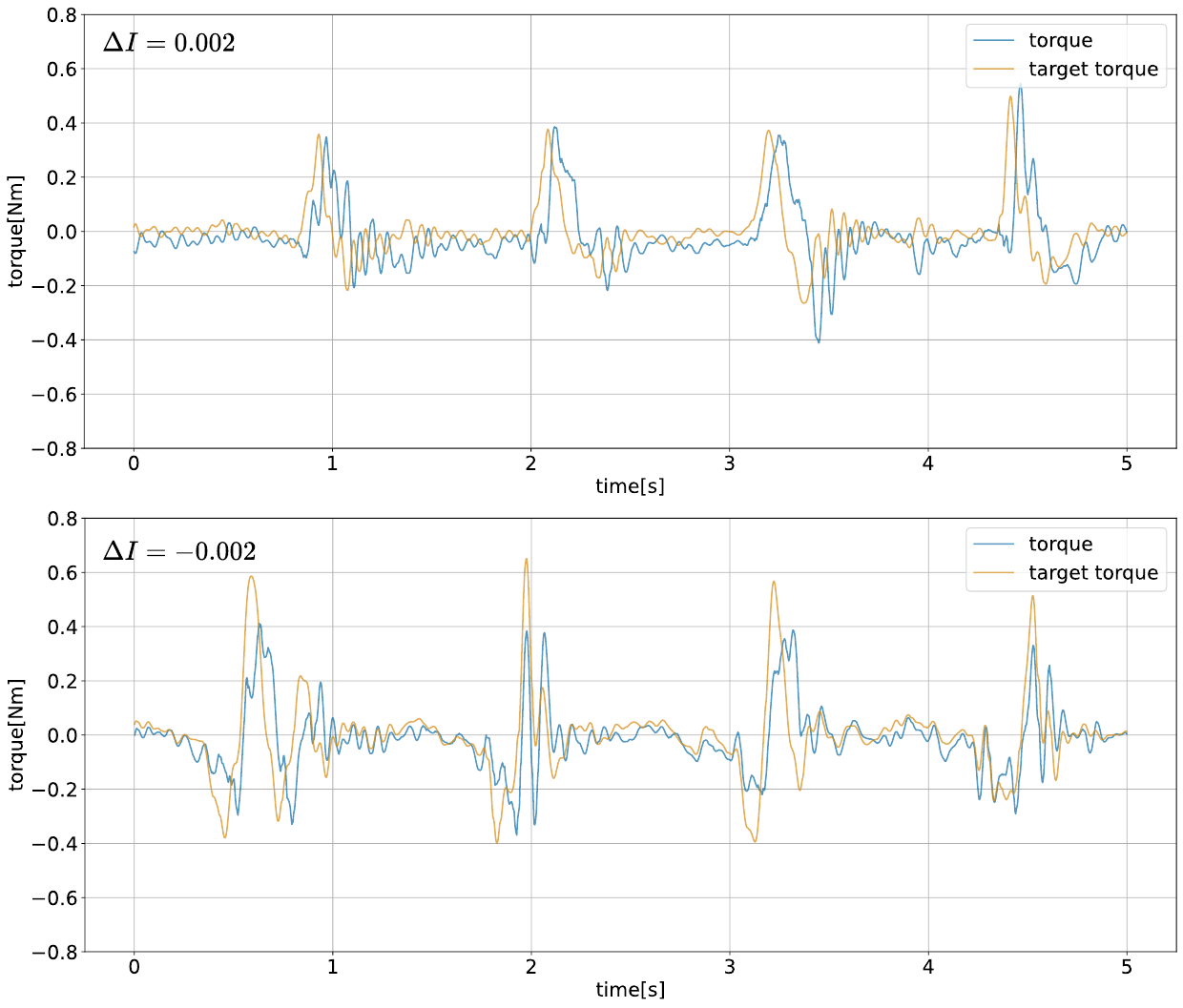}
    \caption{The measured torque and the desired torque for $\Delta I = 0.002, -0.002$.}
    \label{fig:di}
\end{figure}

Figure~\ref{fig:di} shows the temporal variations of the desired torque and the actual output torque when the virtual moment of inertia is increased or decreased.
In a single swinging motion, the system accelerates at the beginning and decelerates when it comes to a stop. 
Thus, the desired torque exhibits a waveform resembling one period of a sine wave with changes from positive to negative or negative to positive within a single swinging motion. 
These sections represent the swinging motion, and it is evident that the actual torque generally follows the desired torque during those periods.

\begin{figure}[]
    \centering
    \includegraphics[width=\linewidth]{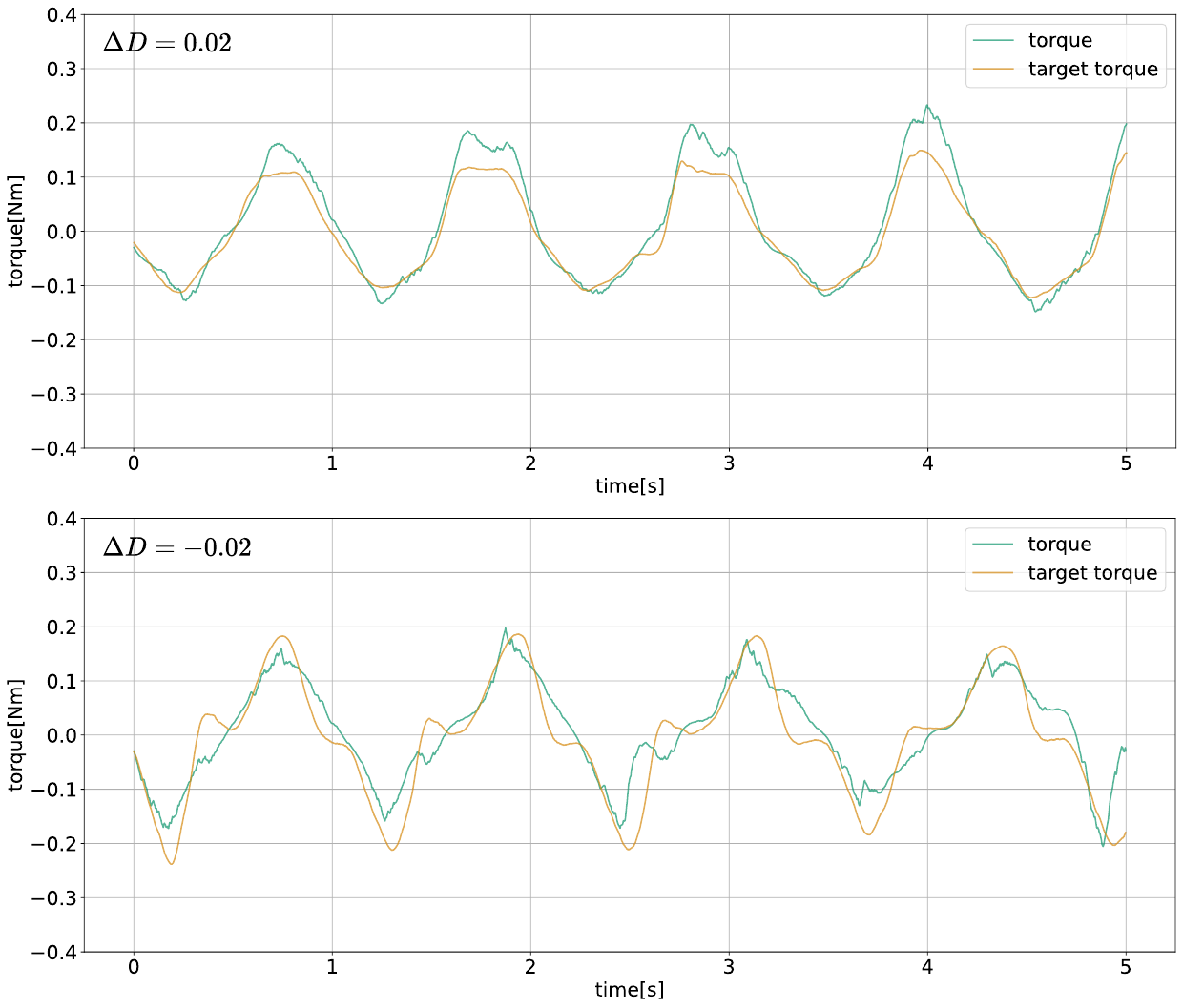}
    \caption{The measured torque and the desired torque for $\Delta D = 0.02, -0.02$.}
    \label{fig:dd}
\end{figure}
\begin{figure}[]
    \centering
    \includegraphics[width=\linewidth]{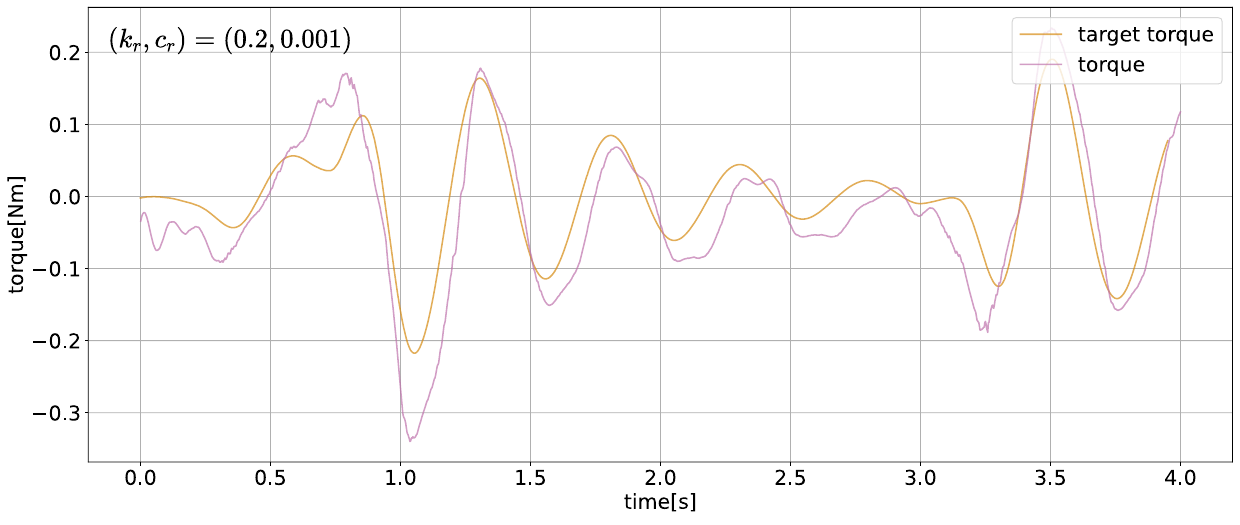}
    \caption{The measured torque and the desired torque for $(k_r, c_r) = (0.2,0.001)$.}
    \label{fig:k-4}
\end{figure}

Figure~\ref{fig:dd} shows the temporal variations of the desired torque and the actual output torque when the virtual viscous damping is varied. 
In the conditions with positive changes in viscous damping, it can be observed that the actual torque closely tracks the desired torque. 
Moreover, for conditions with viscous damping changes of $\Delta D = -0.02$, the system successfully produces the desired torque for swinging motion. 
In the original motion system, viscous damping, a force proportional to velocity, is nearly negligible. Consequently, the overall viscous damping of the motion system represented by $\Delta D < 0$ becomes negative, enabling the system to represent behaviors that are rarely observed in everyday life.

Figure~\ref{fig:k-4} shows the temporal variations of the desired torque and the actual output torque when the virtual stiffness is varied.
The graph shows that oscillatory torque is generated during a single swinging motion.
Note that the swing motion occurs at $t=0.5$.
In this condition, the oscillation converges in approximately 2 seconds within the range of $t=0.5$ to $3.0$, resulting in a frequency of approximately 1/0.5 = 2 Hz.
In summary, in each of the five conditions tested, it is evident that the system consistently and accurately tracks the desired torque.

\section{Interview for haptic perception}\label{sec:interview}
\subsection{Purpose of Interview Study}
In a prior study, we used virtual reality to show how impedance control, which uses force feedback, alters the perception of inertia and resistance during swinging~\cite{Hashimoto2022-af}.
Feedback from user descriptions and conference demos revealed variations in how people respond to the same force feedback. 
These disparities may arise from a fundamental mismatch between real-world physics and sensory perception. 
Consequently, our study delves into how different physical parameters affect human perception.

We intend to employ factor analysis using the semantic differential (SD) method to identify important factors in perceptual experience like prior study \cite{Shirado2005-qk}.
However, the vocabulary used in the SD method should be chosen carefully, as it can only capture a limited number of aspects and variations.
As a result, it is essential to first explore the vocabulary used to describe these experiences.

The purpose of this interview is to explores how users perceive alterations in properties when employing a haptic device. Our emphasis lies in understanding the \textit{how} of this perception rather than the extent, which we aim to comprehend through user interviews. 
Prior to conducting the interview, approval was obtained from the ethics committee of the first author's institution.
\subsection{Interview Design}
In order to investigate what kind of experiences can be imagined through different haptic stimuli created by the force-feedback device, we interviewed three researchers specializing in haptic research, assuming they have a rich imagination linked to haptic sensations.
We asked each researcher to experience the five different haptic stimuli and describe as many phenomena as they could imagine arising from them. 
Additionally, to examine whether the haptic information could be attributed to the handheld tool, we conducted interviews on whether the sensations changed when holding something versus not.

\subsubsection{Selection of Participants}
We conducted interview surveys with three researchers engaged in haptic research because they were expected to have a rich vocabulary and diverse recall of haptic phenomena: P1, an assistant professor at a university (33 years old); P2, a research scientist at a company (31 years old); and P3, another research scientist at a company (34 years old). 
Note that P3 is the second author of this paper.

\subsubsection{Presentation of Stimuli}
We presented five haptic stimuli to the interviewees: ``increased inertia, decreased inertia, increased damping, decreased damping, and decreased stiffness.''
The variables for impedance under each condition were set as shown in Table~\ref{tab:parateters}.
Each participant experienced the stimuli while blindfolded to minimize the influence of visual information. We only instructed them on the direction (pitch direction) of hand movement without specifying the speed or manner. They were advised to improvise the movement based on the sensation of the stimuli.

\subsubsection{Procedure}
As participants experienced each stimulus, we encouraged them to describe as many phenomena and associations as possible. We also inquired whether their sensations changed when gripping a handle or not gripping anything. Participants spent approximately five minutes experiencing and verbalizing their thoughts for each stimulus. The total time for the experiment and interview was around 30 minutes.

\subsection{Thematic Analysis}
We performed thematic analysis on the transcripts created from the recorded interview data~\cite{braun2006using}.
First, the author read the transcripts, divided them into semantic units (codes), and grouped units with the same label. 
From the smallest codes, common abstract codes were extracted, and new codes were created. 
This process was repeated until no commonalities among the higher-level codes were found. 
The resulting higher-level codes were identified as themes.

\subsection{Themes Obtained for Force Feedback}
In the following, we discuss the two main themes identified:  ``Changes in the physical properties of handheld object'' and ``Changes related to bodily sensations'' while citing statements made during the interviews. 
Statements are followed by (participant ID, condition) to clarify which participant made the statement under which force feedback condition.

\subsubsection{Changes in the physical properties of a handheld object}

The changes in the physical properties of held objects extracted from statements can be divided into four categories: rigidity, shape, weight, and viscosity. 

\paragraph{Rigidity}
Under conditions of reduced rigidity, the virtual weight attached to the tip lags behind the motion displacement, exhibiting a swaying behavior when brought to a halt. Statements regarding the rigidity of held objects, such as feeling pulled, a sense of delayed response, swaying, a yo-yo sensation, and a rubber-like displacement, were obtained in response to reduced rigidity conditions.
P3 expressed this sensation as follows:
\begin{quote}\textit{
``After moving, there's a slight delay, and then, the force feedback arrives, as if it's lagging. Maybe the timing isn't quite spot-on. There's a feeling of swaying after I move like the force is being applied later.''} (P3, Reduced Rigidity)
\end{quote}

Additionally, P1 described the swaying and delayed phenomenon as resembling yo-yo motion.

\begin{quote}\textit{
``It's like I'm playing with a yo-yo. But, it's not like I'm swinging it downward, but more in a forward direction. I like the fact that it feels like a yo-yo because I can feel the force after swinging it, like it is being pulled forward.''} (P1, Reduced Rigidity)
\end{quote}

From expressions like ``being pulled'' or ``lagging behind,'' it can be inferred that the participants expected instantaneous force feedback in response to their movements but instead experienced delayed feedback, perceiving a structure that does not immediately follow their motion between their hands and the held object's tip.

\paragraph{Shape}

Many descriptions of the held object's shape changing were observed.
P3 reported feeling as if they were fanning under increased damping conditions and also associated it with holding something wide.
\begin{quote}\textit{
``If I had to describe it in some way, this is a fan. Something broad.''} (P3, increased damping)
\end{quote}

When moving a thin, wide object like a folding fan, the viscous resistance component becomes dominant even in the air. This implies that an increase in viscous resistance evokes such a phenomenon in reality, inducing an image of holding a wide-shaped object.
Moreover, P2 envisioned an expanded circular shape under increased damping conditions, which is consistent with P3's impression of a ``board object.''
\begin{quote}\textit{
``I definitely feel like the tool is stretching. In terms of shape, it's a circle.''} (P2, increased damping)
\end{quote}

Conversely, P2 reported that the object felt elongated under reduced inertia conditions.
\begin{quote}\textit{
``With this, it's a stick, like a really long, slender stick. So, it's incredibly strange that it doesn't hit the table.''} (P2, Reduced Inertia)
\end{quote}

Ideally, decreased inertia should result in the object being perceived as shorter. 
Despite this, the object was reported as feeling longer, possibly because the acceleration felt when swinging the object downward was larger than expected.
As a result, participants may have inferred that the object was longer than its actual size, as longer objects would be expected to have a higher acceleration when swung downward.

\paragraph{Weight and Heaviness}

Participants also reported changes in the perceived weight and heaviness of held objects. 
For instance, P3 described an increase in perceived weight during the increased inertia condition:

\begin{quote}\textit{
``It feels a bit heavier. It feels like I'm holding something heavier than before.''} (P3, Increased Inertia)
\end{quote}

Conversely, P3 reported a decrease in perceived weight under the decreased inertia condition:

\begin{quote}\textit{
``When I drop it, it feels incredibly light, but when I lift it, it feels somewhat heavy. (Omitted) It should be heavy, but it feels light. It's like I can't feel the weight of this device, which gives it a very peculiar sensation.''} (P3, Decreased Inertia)
\end{quote}

It is interesting to note that participants perceived changes in the object's weight even though the object's mass remained constant. This may be due to the changes in the moment of inertia affecting the force feedback and participants interpreting this as a change in weight.

\paragraph{Viscosity}
Participants sometimes attributed the force feedback not to the properties of the held object but to environmental factors.
In particular, descriptions related to viscosity were frequently mentioned under the increased damping condition.

\begin{quote}\textit{
``It feels like I'm in mud. I feel a strong resistance. The movement is delayed. It feels slimy.''} (P2, increased damping)
\end{quote}

Moreover, P1 used the term ``viscosity'' in both the increased inertia and increased damping conditions. While they recognized differences between the two experiences, they still described both using the term ``viscosity.'' This demonstrates that phenomena that can be distinguished at the sensory level can sometimes be represented with similar expressions at the conceptual level.

\begin{quote}\textit{
``It feels like air rather than water. The viscosity has changed. And it feels like there's air resistance or something like that. And also, it feels like my hand has become heavier.''} (P1, Increased Inertia)
\end{quote}
\begin{quote}\textit{
``And this one is more like being underwater than the previous one. The previous one felt like air, but this one has very high viscosity. It feels quite heavy.''} (P1, increased damping)
\end{quote}
\subsubsection{Changes related to bodily sensations}
\paragraph{Body image and abilities}

Contrary to the author's expectations, descriptions were found not only about the handheld tools but also about changes in their body image.
P1 reported a sensation of changes in the shape of their fingers and entire hands. It is uncertain whether the increase in viscosity, for example, can be attributed to the viscosity of the environment or to the properties of one's own body.

\begin{quote}\textit{
``When I tried to swing, I suddenly felt the torque, and I also thought that there was a feeling that my fingers were stretching. 
I think this is largely due to the fact that it becomes difficult to understand the position of my fingers and hands, and I think it is related to that. In the end, I felt that the delayed torque force was stretching my body.''} (P1, reduced stiffness)
\end{quote}

\begin{quote}\textit{
``It feels like there's something like cloth on my hand. It feels like I have fins attached and receiving resistance and swelling up.''} (P1, increased damping)
\end{quote}

Moreover, P2 imagined that their physical abilities had improved. The acceleration of their movement due to decreased inertia can be attributed to enhancing their physical abilities.

\begin{quote}\textit{
``It feels light because of the feedback. And I feel like my strength has increased a lot. I'm not doing it now, but if I were to push something, I feel like I could push it really hard.''} (P2, inertia reduction)
\end{quote}

\begin{quote}\textit{
``Also, I feel my body light. It's more like my body is light rather than what I'm holding.''} (P3, viscosity reduction)
\end{quote}

\paragraph{Body position}

In addition to changes in body shape, there were reports of feeling as if the position of their body was misaligned. This was often seen during the increased damping condition, where the force feedback from moving their hand was greater than expected, causing a sensation of delayed force, and the temporal misalignment was interpreted as spatial misalignment. It is possible that the reverse phenomenon of perceiving force by inserting temporal or spatial delays into movement is occurring and that the occurrence of force is causing the perception of spatial or temporal misalignment.

\begin{quote}\textit{
``I feel like my hand is growing or the position of my hand is somehow misaligned. That's probably because the movement is delayed by one tempo, so the body image becomes misaligned, I think.''} (P2, increased damping)
\end{quote}

\begin{quote}\textit{
``I don't necessarily feel like my hand is in the place where this resistance is felt, but it's true that it's a bit out-of-body-like, so I think it's interesting to create such an effect.
I think the influence of the vibration is quite significant, and because of that, my sense of position around my hand becomes a bit vague.''} (P1, increased damping)
\end{quote}

\paragraph{Sense of agency}

There were also many descriptions regarding the sense of agency, whether they were initiating the movement or not.
P1 mentioned that although the decision to initiate the movement was their own intention, they did not feel like they were driving the movement afterward. This can be attributed to the fact that the feedback proportional to the angular acceleration causes the entire movement system to move at a speed beyond their prediction, making them feel as if they cannot control it.
\begin{quote}\textit{
``There is a slight sense of incongruity in accelerating, not like being pushed a little, but more like a clear sense of being moved. It's that strong. 
Although I'm only pulling the initial trigger, I don't have much awareness that I'm driving the movement afterward. 
Also, I'm not really conscious of stopping it.''} (P1, inertia reduction)
\end{quote}
On the other hand, P1 described a sense of being able to control the movement under the viscosity reduction condition.
\begin{quote}\textit{
``If I want to stop, it stops there, or rather, it feels like it won't accelerate further, so there is a sense of agency. I can tell it's my movement.''} (P1, viscosity reduction)
\end{quote}

Moreover, they mentioned that after a few swings, their prediction model changes, and they feel like they can control the movement. This suggests that feedback proportional to velocity may be easier to adapt to than feedback proportional to acceleration.

\begin{quote}\textit{
``As I get used to it, I feel like I can stop it where I want to stop. It takes some effort, but the response is better than before (inertia reduction).''} (P1, viscosity reduction)
\end{quote}

\subsection{Discussion}
Surpassing our expectations, we discovered that various phenomena could be evoked through haptic stimulation.
The results regarding the user's perception of object properties and interactions with the environment were largely as anticipated. 
When force was exerted in the direction of increasing inertia, objects became more tough to move, their lengths seemed longer, and the sensation of holding a larger object arose. 
Similarly, when force was applied in the direction of increasing viscosity, objects became more tough to move, and a more viscous, sticky sensation was often reported. 
However, these ``difficult to move'' sensations were frequently confounded, with reports of increased viscosity under increased inertia or increased object size under conditions of increased viscosity. 
Further investigation through quantitative experiments is necessary to determine how much each force presentation influences human representations, such as ``size'' and ``stickiness.''

Moreover, it was found that under conditions of reduced rigidity, expressions of ``flexibility'' and ``springiness'' were possible. 
It was also revealed that under conditions of decreased rigidity, phenomena were more difficult to associate with something specific than conditions of inertia or viscosity changes. 
This difficulty might be attributed to predicting motion based solely on force feedback. In the other four conditions, there was feedback on velocity and acceleration at each point in time, resulting in high reproducibility of movement. 
In contrast, conditions of reduced rigidity involved simulating specific object movements, which led to more complex behavior and made it harder to imagine the nature of the motion. 
This difficulty in imaging and predicting motion can be inferred from the tendency to repeat swinging motions multiple times before verbalizing the phenomenon under conditions of decreased rigidity. On the other hand, the diversity of experienced images will change according to the given language. 
It may be necessary to investigate how various motion-describing vocabularies relate to different representations.

In conditions of accelerating motion with decreased inertia and viscosity, there were fewer opinions attributing the experience to the properties of the handheld object, with more reports attributing the experience to the transformation of one's own body.
This may be due to the behavior of objects appearing to have negative inertia when attempting to express smaller inertia than the handheld object's inherent inertia. 
In reality, objects that behave as if they have negative inertia (starting to move on their own when trying to accelerate) do not exist. 
To interpret such unlikely phenomena, the possibility of attributing them to changes in one's body's properties (e.g., lighter hands, stronger force) rather than the tool itself may arise. 
Additionally, it was found that the experience of bodily transformation is more likely to be attributed to a relaxed state than holding or grasping an object. 

\subsection{Conclusion of Interview Study}
In this section, researchers experienced various haptic stimuli to investigate what impressions force feedback could evoke regarding handheld objects. As a result, it was confirmed that perceptions of size, weight, and environmental viscosity change depending on the type of haptic stimulation. 
From these results, it was determined that to attribute the information from each force feedback to the physical properties of the handheld object, it is advisable to perform movements while the user imagines holding something and to consider the device's inherent physical properties.
In the following section, based on the various impressions evoked by the force feedback obtained from this interview, an investigation will be conducted to determine how much each force feedback technique influences each impression.

\section{Evaluation of haptic perception and impedance change}
In the previous section, a qualitative investigation explored the potential associations between varying types of force feedback and their resulting perceptions. This allowed for a certain degree of prediction regarding the perceptions evoked by the proposed device's motion intervention experiences. However, the extent to which each perception is linked to changes in individual impedance cannot be determined solely through qualitative research.

Therefore, this section aims to clarify the necessary parameter adjustments to evoke each perception by investigating the degree to which individual impedance changes affect the various perceptions obtained through qualitative research using experiments with human participants.
As an experimental method to identify the main factors contributing to perceptual experiences, the previous section mentioned factor analysis on data obtained through the ``semantic differential'' method (SD method), which involves evaluating a single stimulus using various adjective pairs. 
The advantage of this method lies in its ability to extract dimensions of the space representing human dynamic touch perceptions of physical properties (material sensations) through multivariate analysis. 

\subsection{Selection of Word Pairs for Property Perception Evaluation}\label{sec:words}

In this section, we select adjective pairs for evaluation in the experiment.
We selected adjectives corresponding to themes and categories obtained from the interviews in the section \ref{sec:interview}.
Since it may be difficult to describe some types of force feedback using adjectives, we also selected words that include verbs and nouns frequently used by participants for such stimuli.
In addition, we selected words from the obtained themes related to the sensations of the handheld object itself and the interaction between the handheld object and the environment, such as ``changes in the physical properties of the held tool,'' ``interactions occurring between the environment,'' and ``changes in motion when moving the arm.''

In word selection, it is crucial to strike a balance.
We aim to represent various aspects while keeping the burden on participants low and avoiding word interactions. 
We chose 8 words with low similarity.
Here are the selected words: Long -- Short, Wide -- Narrow, Thick -- Thin, Hard -- Soft, Heavy -- Light, Stiff -- Flexible and Sticky -- Smooth.
\subsection{Experiment}
Using the word list selected in the section \ref{sec:words}, we investigated to understand how the proposed device's motion intervention produces various sensations in humans.

\subsubsection{Participants}
The participants were 16 people aged 21 to 24 (1 female, 15 males), with an average age of 22.75 (standard deviation 1.06 years).

\subsubsection{Haptic stimuli used in the experiment}
As in the section~\ref{sec:measurement}, five types of haptic stimuli were prepared. The stimuli were ``increased inertia, decreased inertia, increased damping, decreased damping, and decreased stiffness.''

\subsubsection{Procedure}

\begin{figure}
    \centering
    \includegraphics[width=\linewidth]{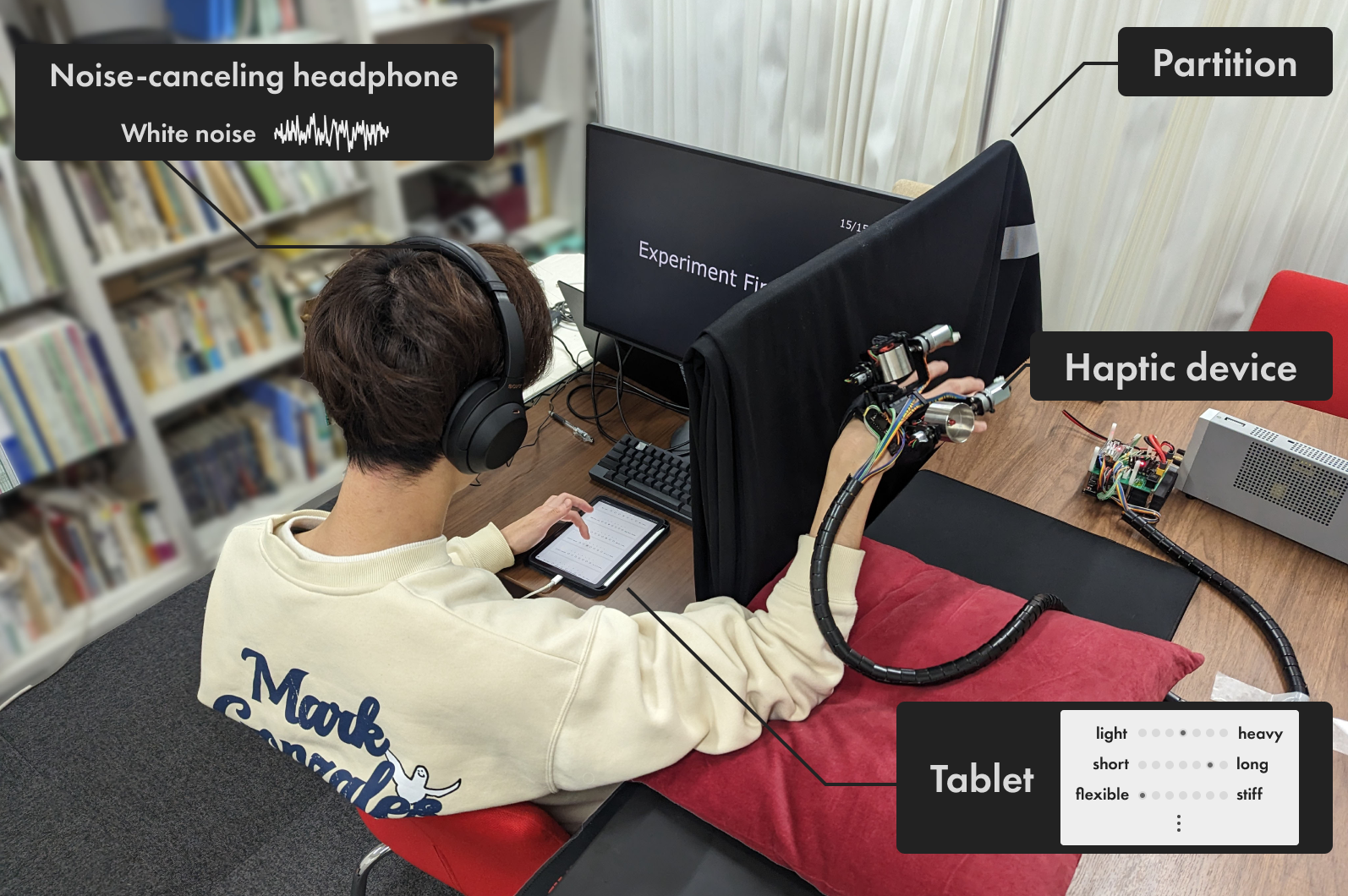}
    \caption{Participants are wearing noise-canceling headphones and a haptic device. The participant evaluates on a tablet while shaking their wrist.}
    \label{fig:experiment}
\end{figure}

Participants wore the haptic device on their right hand and experienced haptic stimuli. 
To minimize the influence of the device's movement and the sound emitted, the participants placed their arms behind a partition, as shown in Figure~\ref{fig:experiment}, and performed swinging motions. In addition, participants wore noise-canceling headphones, and white noise was played to prevent them from hearing the sounds emitted by the device.
Participants could experience force feedback while moving their right hand. 
While performing swinging motions, they evaluated how well the haptic stimuli corresponded to the words selected in the previous section using a tablet. The evaluation was conducted using a 7-point Likert scale. After evaluating all words for one stimulus, the screen switched to a waiting screen.
Participants could rest until their arm fatigue subsided and proceed to the next trial at their own pace. They repeated the evaluation three times for all five experimental stimuli. The experiment took approximately 45 minutes, including the explanation. Participants were rewarded with an Amazon gift card worth \$6 as compensation.
Prior to conducting this experiment, approval was obtained from the ethics committee of the first author's institution.

\subsection{Data Analysis}
Since evaluations for the five types of stimuli were repeated three times for 13 words, we averaged the evaluated values for each stimulus. 
As it is highly likely that the pairs of words are correlated, we attempt to extract lower-dimensional factors through factor analysis.

\begin{figure}
    \centering
    \includegraphics[width=\linewidth]{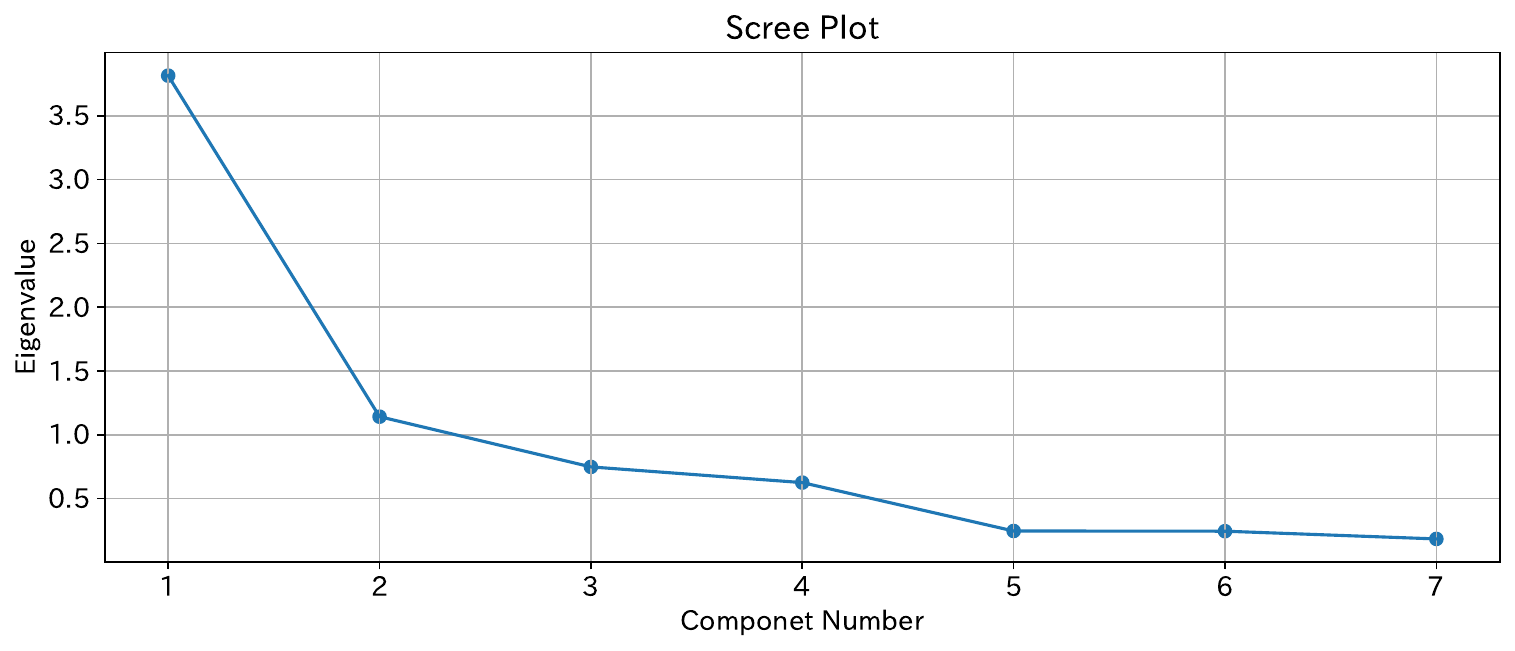}
    \caption{Scree plot depicting eigenvalues for principal components. The plot illustrates the diminishing contribution of each component to the total variance in the dataset.}
    \label{fig:scree_plot}
\end{figure}

\begin{figure}
    \centering \includegraphics[width=\linewidth]{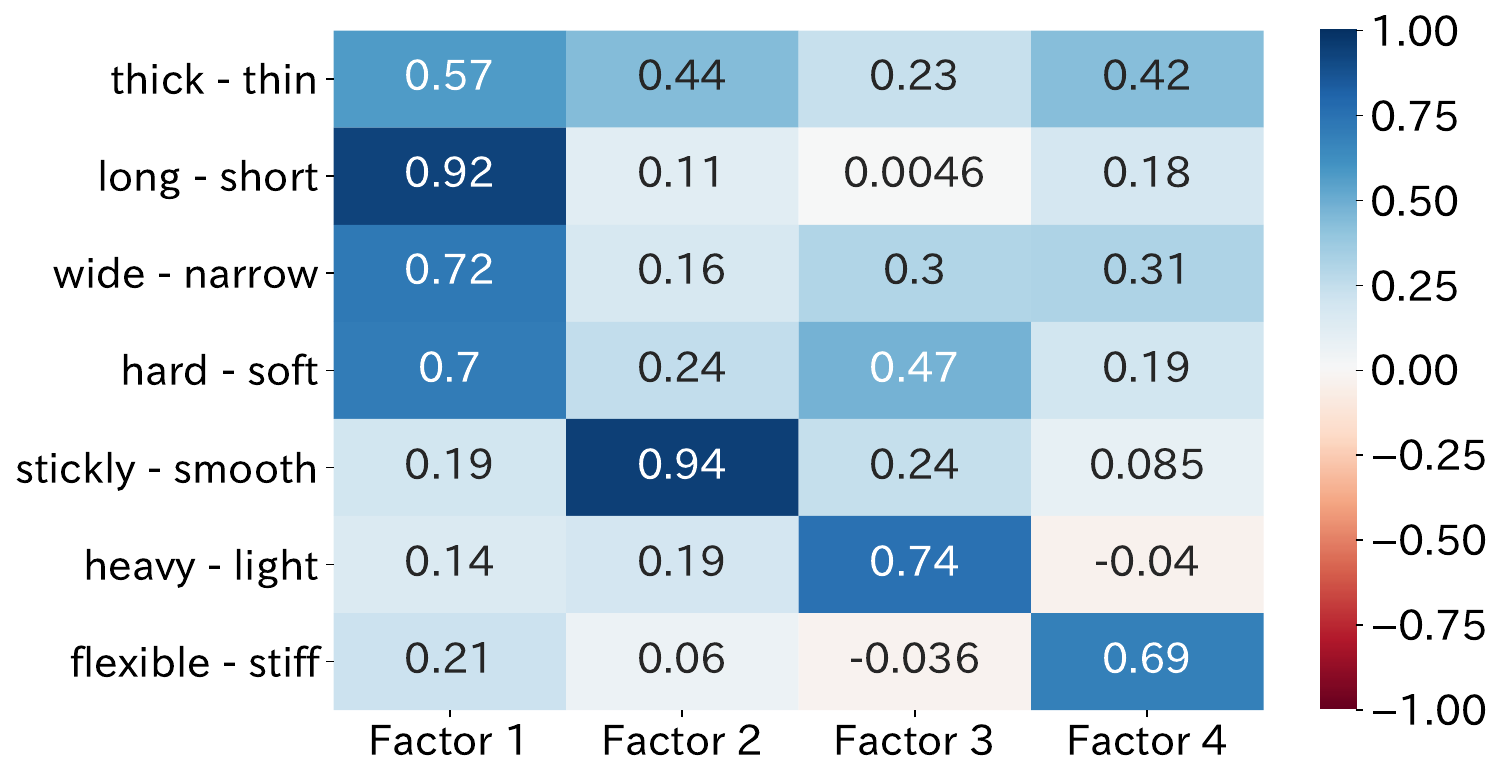}
    \caption{Heatmap of Word-Factor Associations.}
    \label{fig:factor_heatmap}
\end{figure}

Figure~\ref{fig:scree_plot} shows the scree plot of the factor analysis. 
We chose the number of factors as 4, where the second-largest change in the slope of the broken line occurred. 
To make the four factors more comprehensible, we apply varimax rotation to the results.
The heatmap of loadings after the rotation is shown in Figure~\ref{fig:factor_heatmap}. 
The squared factor loadings, the contribution rates of each factor, and the cumulative contribution rates are shown in Table~\ref{tab:factor_info}. 
The four factors can explain 75.8~\% of the data.
In studies on the extraction of haptic dimensions, factors and their cumulative contribution rates are often around 79~\%~\cite{Shirado2004KJ00007553977}.
Compared to those studies, the contribution rate by factors in this study is slightly lower. Several causes can be considered for this.
First, in the extraction of haptic dimensions using conventional haptic stimuli, evaluations are conducted on existing objects, so the participants are more likely to have touched those objects before. 
In contrast, in this study, the expressions of simulated material properties are conveyed through a haptic device, making it challenging to represent the complete information as when swinging real objects. 
Additionally, some experiences were not grounded in words, such as negative inertia and negative viscosity, which cannot be experienced with objects in the real world. 
Therefore, individual differences among participants in attributing these experiences to specific material properties may have increased, leading to a lower contribution rate explained by the factors.

\begin{table}
    \centering
    \caption{Sum of Squared Loadings and the factor's contribution}
    \resizebox{\linewidth}{!}{
    \begin{tabular}{lrrrr}
    \toprule
    {} &  Factor 1 &  Factor 2 &  Factor 3 &  Factor 4 \\
    \midrule
    Sum of Squared Loadings &  2.283029 &  1.221785 &  0.982619 &  0.821623 \\
    \% of Variance       &  0.326147 &  0.174541 &  0.140374 &  0.117375 \\
    Cumulative \%     &  0.326147 &  0.500688 &  0.641062 &  0.758437 \\
    \bottomrule
    \end{tabular}}
    \label{tab:factor_info}
\end{table}

\begin{figure}
    \centering \includegraphics[width=\linewidth]{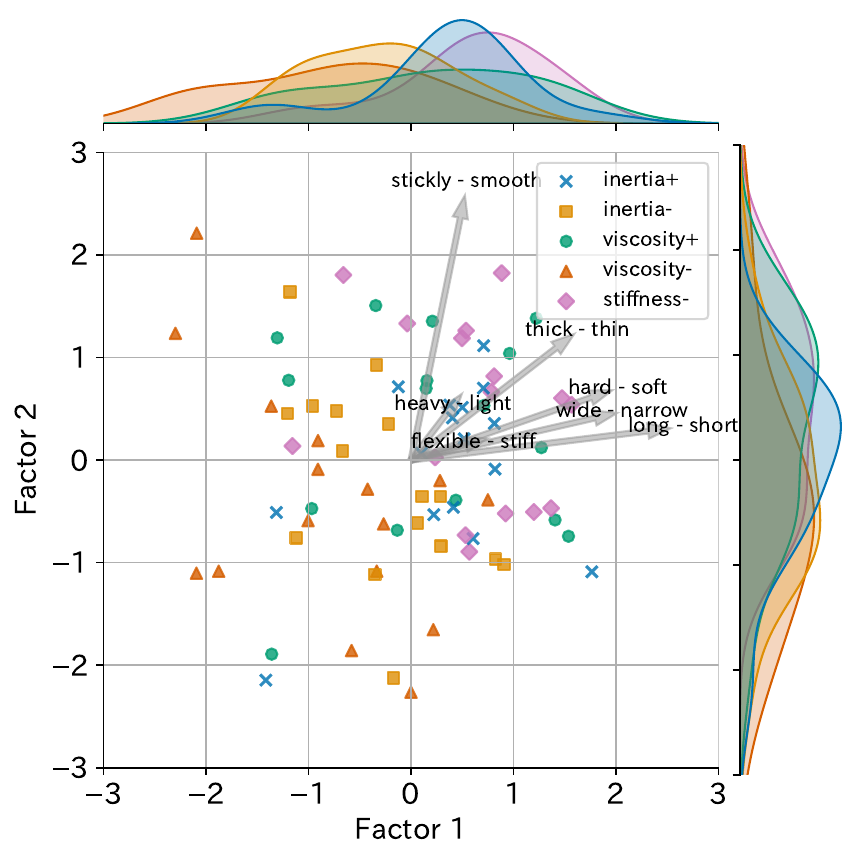}
    \caption{bi-plot between factor 1 and factor 2 dimension}
    \label{fig:factor1_and_factor2_biplot_joint}

    \centering \includegraphics[width=\linewidth]{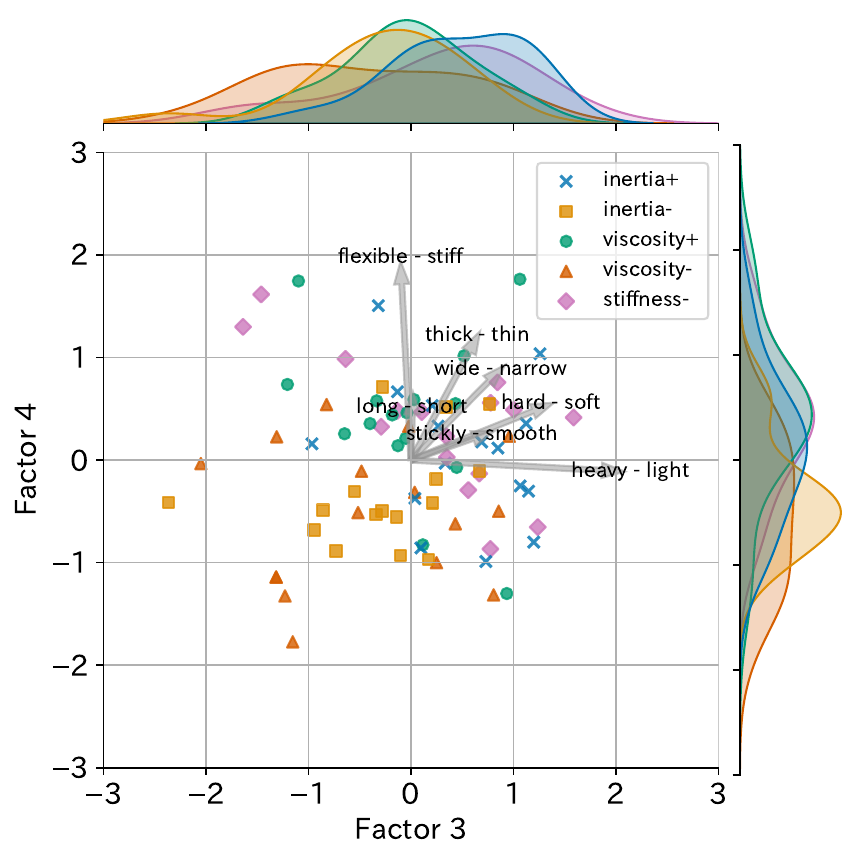}
    \caption{bi-plot between factor 3 and factor 4 dimension}
    \label{fig:factor3_and_factor4_biplot_joint}
\end{figure}

Figure~\ref{fig:factor1_and_factor2_biplot_joint} and \ref{fig:factor3_and_factor4_biplot_joint} show the arrangement of each stimulus in the factor score space. A small marker represents each data point, and each arrow indicates what words are associated with each factor.

\subsection{Discussion}
\subsubsection{Property perception dimensions}
From the results of the factor analysis, it is considered that the factors of the material perception of the handheld object by the proposed method are mainly: ``size factor'', ``viscosity factor'', ``weight factor'', and ``flexibility factor''.

For the first factor, ``size factor'', it can be seen that words such as ``long'', ``thick'', ``wide'', and ``hard'' are well connected. 
This factor can be inferred to be strongly associated with objects like long, thick, and rigid cubes, like a rod. 
In this experiment, the participants were asked to perform pitch-swinging motions. 
As revealed in experiments by Turvey~\cite{Turvey1995-kn} and Zenner~\cite{Zenner_2017}, the perception of an object's size is influenced by the principal moment of inertia around the rotation axis. 
The object's length and thickness determine the principal moment of inertia in the pitch direction and is not much related to width $(I = \frac{1}{3}m(l^2 + d^2))$. 
Nevertheless, the inclusion of the component ``wide'' may be influenced by the factor of spatial spread, such as length and thickness.

The second factor, ``viscosity factor'', was strongly associated with words such as ``sticky'' and ``thick''. As hypothesized, the viscosity factor in the environment appeared due to the force proportional to velocity. 
However, it was unexpected that it was associated with ``thickness,'' which is the spatial spread of the object. Instead, if one were to predict the spatial spread of the object from the viscosity factor, one would imagine a thin object like a fan so that the thinness factor would become stronger. Contrary to this, it is suggested that users imagined a thick object in a high-viscosity environment. By the way, the feeling of stickiness was expressed as Sticky--Smooth, and thickness as Thick-Thin. However, the term ``thick liquid'' is sometimes used to describe a high-viscosity liquid. The word ``thick'', describing the density of the liquid, and the word ``thick'', describing the spatial spread, may be connected in the brain, possibly drawing out the correlation between the two in a top-down manner.

The third factor, ``weight factor'', had a strong association with words such as ``heavy'' and ``hard''. While it may seem that spatial spread and the weight of an object are, of course, correlated in reality (indeed, larger objects are more likely to be heavier), this analysis resulted in a separate independent factor. 
This is also consistent with Turvey et al.'s claim that the weight of an object does not affect the perception of spatial spread through dynamic touch~\cite{Turvey1998}. 
However, it should be noted that the word ``heavy'' includes not only the properties of the object but also the lightness and heaviness of one's movement, so it may not purely describe the weight of the object.

The fourth factor, ``flexibility factor'', had a strong association with words such as ``flexible''. This corresponds to the flexibility of the object. Also, this flexibility factor has a weak correlation with words such as thickness and width. 
In terms of thickness, it may be since one cannot feel something like bending unless the object has a certain thickness. Regarding width, there were reports that people imagined objects with flexibility, like fans or placemats, when experiencing flexibility. 
There is also a similar term, ``hardness,'' but the impact of this term on the flexibility factor was small. This is probably due to the distinction between the flexibility and softness of the object. 
Flexible objects are those with a core and do not sag under their weight when held, while soft objects easily deform under their weight.

\begin{figure}
    \centering
    \includegraphics[width=\linewidth]{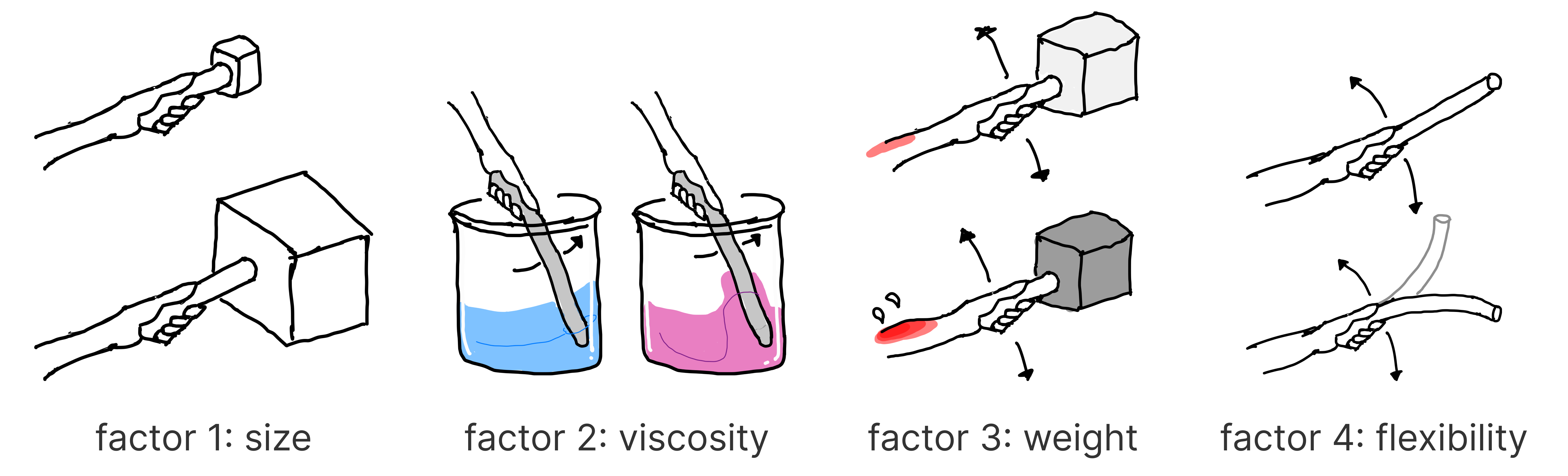}
    \caption{Examples of object properties and environmental properties for each factor.}
    \label{fig:factor_visualization}
\end{figure}

Examples of object properties and environmental properties that change with these factors are shown in Figure \ref{fig:factor_visualization}.
For the spatial spread factor, the size of the object changes.
For the viscosity factor, the difference in viscosity in the environment when swinging the object is represented.
From the weight factor, the weight of the object itself and the difficulty of motion change.
From the flexibility factor, one can imagine objects made of rubber-like materials that are flexible to hard objects.
\subsubsection{Relationship between impedance and property factors}
\begin{figure}
    \centering
    \includegraphics[width=\linewidth]{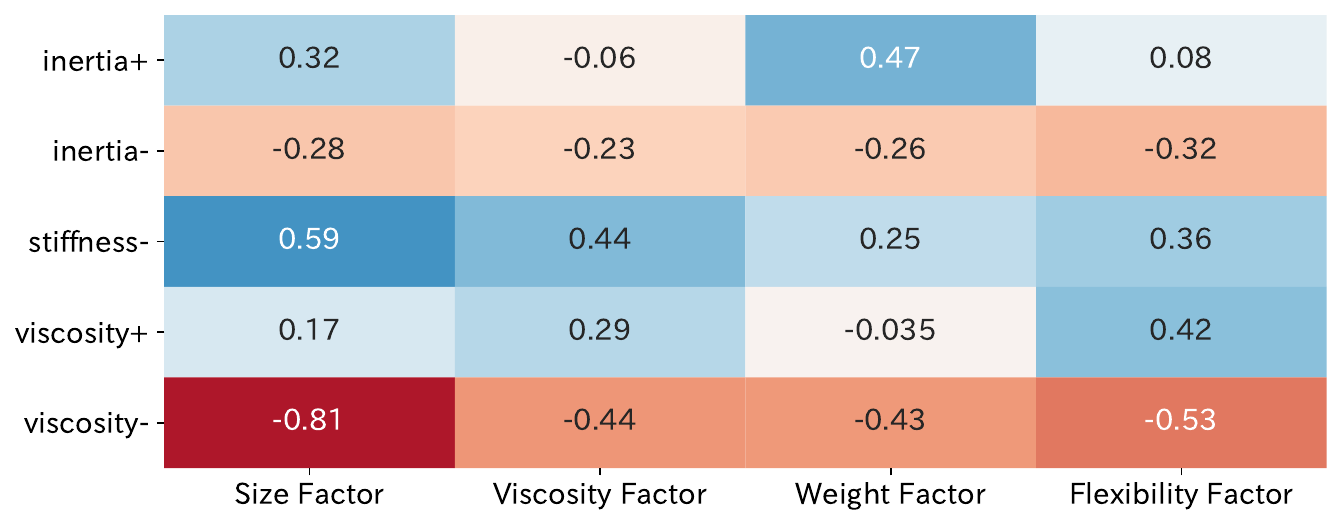}
    \caption{Mean of factor scores in each impedance change.}
    \label{fig:heatmap}
\end{figure}

Now that we have extracted the factors by motion intervention, let us see how the impedance is related to each factor.
Figure~\ref{fig:heatmap} shows the mean of factor scores calculated for each impedance change. 

In the increased inertia condition, an increase in size factor and weight factor is observed. 
This shows that the increase in the moment of inertia in the motion enlarges the perception of spatial spread and further increases the perception of weight. 
Also, it does not have much effect on viscosity and flexibility.

In the decreased inertia condition, there is a tendency for all factors to decrease moderately overall. 
It is assumed that the perception of size, viscosity, and weight decreased as the motion accelerated. It is also related to the image of being rigid, which may be due to the agility of the movement. 
The motion of a flexible object gives the impression of being vibrational and slow-moving. In contrast, the motion of a rigid object gives the impression of being linear and fast-moving, which is thought to be related to rigidity.

In the increased damping condition, in contrast to the increased inertia condition, it is found to affect Factor 2 and Factor 4, stickiness, and flexibility. The perception of stickiness increases as the increased dampings. In addition, there is a positive correlation with the perception of flexibility. It was a surprising result that flexibility was also associated with increased viscosity. One possible reason for this is that the words flexibility were projected as a result of trying to attribute the force, like viscosity during motion, to the properties of the tool. It is known that objects with high flexibility are more affected by viscous resistance than those with low flexibility (results of multi-particle simulations). A high viscous resistance may lead the user to infer that the object is flexible.

In the decreased viscosity condition, a similar tendency to the decreased inertia condition was observed, but the extent was greater than that of the decreased inertia. This is because the feedback in the direction of accelerating the movement causes acceleration even with a slight movement by the user, and the sensation of being accelerated is stronger than during the decreased inertia. It is known that the sensation of moving by oneself is smaller during the decreased viscosity. As a result, the perception of size, viscosity, and weight has decreased. Regarding rigidity, as in the case of decreased inertia, rigidity is imagined from the speed of movement.

For the decreased stiffness condition, a positive correlation was observed for all factors. Regarding the size factors, the score is somewhat high due to the prediction that the object must be somewhat long for elastic motion to occur. It would be difficult to feel the flexibility when shaking a short object. 
Also, regarding viscosity, as in the case of increased damping conditions, the flexibility of the object is more likely to occur in a high-viscosity environment. The magnitude of viscosity and the magnitude of flexibility can be perceived as linked. 
Lastly, regarding weight, the force stimulus is delayed in the initial movement under the increased elasticity condition compared to other conditions. This temporal delay is thought to evoke the image of an object with a weight at the tip, increasing the perception of the object's weight.

It is also worth mentioning the differences between the conditions. 
Both the increased inertia condition and the increased damping condition produce forces that hinder movement. 
In the experiment, there were reports that it was difficult to express the difference between these two in words, but they could be understood. 
However, the results showed that the increased inertia affected the perception of the object's weight and size. In contrast, the increased damping condition affected the environmental viscosity and flexibility, resulting in independent effects. This indicates that people can not only distinguish these two stimuli but also treat them separately at the layer of meaning attribution.

We must also mention the non-linearity between the impedance change and the factors.
The size and weight factors increased under conditions of increased inertia, whereas all four factors tended to decrease under conditions of decreased inertia. 
Similarly, the flexibility and viscosity factors increased under increased damping, while all factors tended to decrease under decreased viscosity.
We suspect this nonlinear relationship for the same impedance is due to the rarity of sensory information.
The experience of increased inertia and viscosity can be felt in everyday life by holding and shaking a tool or stirring water with one's hands.
Therefore, it is easy to attribute the increase in inertia or viscosity to a specific factors such as the length of an object or a sticky liquid. 
On the other hand, we rarely experience negative values of inertia and viscosity in the real world.
Therefore, it is not tied to a specific physical property, but could be attributed to any of the four properties extracted, such as the shortness or lightness or rigidity of the object, or low viscosity in the medium.
This is not determined by haptic sensation alone, but depends on what is given as visual information.

\section{Conclusion}
The objective of this paper was to elucidate the dimensions present in the haptic experiences that alter with the impedance of motion and to investigate the extent of influence each impedance alteration exerts on these dimensions. 
Through evaluations employing pairs of phrases, we derived four dimensions:
size (perception related to the size and shape of an object)
weight (perception of an object's weight)
viscosity (perception of the viscosity around an environment)
flexibility (perception of an object's flexibility) 
Furthermore, we clarified the extent of influence each impedance change has on these four dimensions.

However, it is essential to note that more than this paper is needed to provide a complete understanding of how perceptions might manifest when various methods are combined. 
Whether the combination of these perceptions can effectively express each dimension requires further investigation.

The accomplishments obtained from this experiment imply that changes in each impedance do not independently trigger alterations in perception but rather invoke various phenomena. 
This poses a challenge that conventional psychophysical experiments, which often focus on averages, cannot address. Particularly in the realm of haptic phenomena, where diversity in perception is expected, it is essential to engage in not only discussions of classical averages but also qualitative research to discern this diversity.

With these findings in mind, future research in the haptics field should undertake a more detailed analysis of the impact of impedance changes on perceptual dimensions and explore the interactions between these dimensions. 
Furthermore, conducting verification in real-world scenarios, such as incorporating visual stimuli, to establish the relationship between impedance changes and perceptual dimensions and subsequently connecting this to practical applications is an anticipation for future study.

\bibliographystyle{IEEEtran}
\bibliography{ref.bib}

\begin{thebibliography}{10}
\providecommand{\url}[1]{#1}
\csname url@samestyle\endcsname
\providecommand{\newblock}{\relax}
\providecommand{\bibinfo}[2]{#2}
\providecommand{\BIBentrySTDinterwordspacing}{\spaceskip=0pt\relax}
\providecommand{\BIBentryALTinterwordstretchfactor}{4}
\providecommand{\BIBentryALTinterwordspacing}{\spaceskip=\fontdimen2\font plus
\BIBentryALTinterwordstretchfactor\fontdimen3\font minus \fontdimen4\font\relax}
\providecommand{\BIBforeignlanguage}[2]{{%
\expandafter\ifx\csname l@#1\endcsname\relax
\typeout{** WARNING: IEEEtran.bst: No hyphenation pattern has been}%
\typeout{** loaded for the language `#1'. Using the pattern for}%
\typeout{** the default language instead.}%
\else
\language=\csname l@#1\endcsname
\fi
#2}}
\providecommand{\BIBdecl}{\relax}
\BIBdecl

\bibitem{Shirado2005-qk}
H.~Shirado and T.~Maeno, ``Modeling of human texture perception for tactile displays and sensors,'' in \emph{First Joint Eurohaptics Conference and Symposium on Haptic Interfaces for Virtual Environment and Teleoperator Systems. World Haptics Conference}, Mar. 2005, pp. 629--630.

\bibitem{Bergmann_Tiest2006-qg}
W.~M. Bergmann~Tiest and A.~M.~L. Kappers, ``\BIBforeignlanguage{en}{Analysis of haptic perception of materials by multidimensional scaling and physical measurements of roughness and compressibility},'' \emph{\BIBforeignlanguage{en}{Acta Psychol.}}, vol. 121, no.~1, pp. 1--20, Jan. 2006.

\bibitem{Turvey1998}
M.~T. Turvey, G.~Burton, E.~L. Amazeen, M.~Butwill, and C.~Carello, ``\BIBforeignlanguage{en}{Perceiving the width and height of a hand-held object by dynamic touch},'' \emph{\BIBforeignlanguage{en}{J. Exp. Psychol. Hum. Percept. Perform.}}, vol.~24, no.~1, pp. 35--48, Feb. 1998.

\bibitem{burton1990can}
G.~Burton, M.~T. Turvey, and H.~Y. Solomon, ``\BIBforeignlanguage{en}{Can shape be perceived by dynamic touch?}'' \emph{\BIBforeignlanguage{en}{Percept. Psychophys.}}, vol.~48, no.~5, pp. 477--487, Nov. 1990.

\bibitem{Turvey1996-ml}
M.~T. Turvey, ``Dynamic touch,'' \emph{Am. Psychol.}, vol.~51, no.~11, pp. 1134--1152, Nov. 1996.

\bibitem{lederman1987hand}
S.~J. Lederman and R.~L. Klatzky, ``Hand movements: A window into haptic object recognition,'' \emph{Cognitive psychology}, vol.~19, no.~3, pp. 342--368, 1987.

\bibitem{Turvey1995-kn}
M.~T. Turvey and C.~Carello, ``Dynamic touch,'' \emph{Perception of space and motion}, 1995.

\bibitem{Hollins1993-eu}
M.~Hollins, R.~Faldowski, S.~Rao, and F.~Young, ``\BIBforeignlanguage{en}{Perceptual dimensions of tactile surface texture: a multidimensional scaling analysis},'' \emph{\BIBforeignlanguage{en}{Percept. Psychophys.}}, vol.~54, no.~6, pp. 697--705, Dec. 1993.

\bibitem{Park2023-kr}
C.~Park, J.~Kim, and S.~Choi, ``Visuo-haptic crossmodal shape perception model for {Shape-Changing} handheld controllers bridged by inertial tensor,'' in \emph{Proceedings of the 2023 {CHI} Conference on Human Factors in Computing Systems}, ser. CHI '23, no. Article 699.\hskip 1em plus 0.5em minus 0.4em\relax New York, NY, USA: Association for Computing Machinery, Apr. 2023, pp. 1--18.

\bibitem{fujinawa2017computational}
E.~Fujinawa, S.~Yoshida, Y.~Koyama, T.~Narumi, T.~Tanikawa, and M.~Hirose, ``{Computational design of hand-held VR controllers using haptic shape illusion},'' in \emph{Proceedings of the 23rd ACM Symposium on Virtual Reality Software and Technology}.\hskip 1em plus 0.5em minus 0.4em\relax ACM, 2017, p.~28.

\bibitem{Zhu2019-aa}
K.~Zhu, T.~Chen, F.~Han, and Y.-S. Wu, ``{HapTwist}: Creating interactive haptic proxies in virtual reality using low-cost twistable artefacts,'' in \emph{Proceedings of the 2019 {CHI} Conference on Human Factors in Computing Systems}, ser. CHI '19, no. Paper 693.\hskip 1em plus 0.5em minus 0.4em\relax New York, NY, USA: Association for Computing Machinery, May 2019, pp. 1--13.

\bibitem{Chen2021-rg}
Y.-W. Chen, W.-J. Lin, Y.~Chen, and L.-P. Cheng, ``{PneuSeries}: {3D} shape forming with modularized {Serial-Connected} inflatables,'' in \emph{The 34th Annual {ACM} Symposium on User Interface Software and Technology}, ser. UIST '21.\hskip 1em plus 0.5em minus 0.4em\relax New York, NY, USA: Association for Computing Machinery, Oct. 2021, pp. 431--440.

\bibitem{9284771}
M.~Feick, S.~Bateman, A.~Tang, A.~Miede, and N.~Marquardt, ``Tangi: Tangible proxies for embodied object exploration and manipulation in virtual reality,'' in \emph{2020 IEEE International Symposium on Mixed and Augmented Reality (ISMAR)}, 2020, pp. 195--206.

\bibitem{Swindells_2003}
C.~Swindells, A.~Unden, and T.~Sang, ``{TorqueBAR}: an ungrounded haptic feedback device,'' in \emph{Proceedings of the 5th international conference on Multimodal interfaces}, ser. ICMI '03.\hskip 1em plus 0.5em minus 0.4em\relax New York, NY, USA: Association for Computing Machinery, Nov. 2003, pp. 52--59.

\bibitem{Zenner_2017}
A.~Zenner and A.~Kr{\"{u}}ger, ``{Shifty: A Weight-Shifting Dynamic Passive Haptic Proxy to Enhance Object Perception in Virtual Reality},'' \emph{IEEE Transactions on Visualization and Computer Graphics}, 2017.

\bibitem{shigeyama2019transcalibur}
J.~Shigeyama, T.~Hashimoto, S.~Yoshida, T.~Narumi, T.~Tanikawa, and M.~Hirose, ``Transcalibur: A weight shifting virtual reality controller for {2D} shape rendering based on computational perception model,'' in \emph{Proceedings of the 2019 {CHI} Conference on Human Factors in Computing Systems}, ser. CHI '19, no. Paper 11.\hskip 1em plus 0.5em minus 0.4em\relax New York, NY, USA: Association for Computing Machinery, May 2019, pp. 1--11.

\bibitem{Gonzalez_Avila2021-ll}
J.~F. Gonzalez~Avila, J.~C. McClelland, R.~J. Teather, P.~Figueroa, and A.~Girouard, ``Adaptic: A shape changing prop with haptic retargeting,'' in \emph{Proceedings of the 2021 {ACM} Symposium on Spatial User Interaction}, ser. SUI '21, no. Article 4.\hskip 1em plus 0.5em minus 0.4em\relax New York, NY, USA: Association for Computing Machinery, Nov. 2021, pp. 1--13.

\bibitem{zenner2019drag}
A.~Zenner and A.~Kr{\"u}ger, ``Drag:on: A virtual reality controller providing haptic feedback based on drag and weight shift,'' in \emph{Proceedings of the 2019 {CHI} Conference on Human Factors in Computing Systems}, ser. CHI '19, no. Paper 211.\hskip 1em plus 0.5em minus 0.4em\relax New York, NY, USA: Association for Computing Machinery, May 2019, pp. 1--12.

\bibitem{10.1145/3305367.3327991}
Y.~Liu, T.~Hashimoto, S.~Yoshida, T.~Narumi, T.~Tanikawa, and M.~Hirose, ``{ShapeSense}: a {2D} shape rendering {VR} device with moving surfaces that controls mass properties and air resistance,'' in \emph{{ACM} {SIGGRAPH} 2019 Emerging Technologies}, ser. SIGGRAPH '19, no. Article 23.\hskip 1em plus 0.5em minus 0.4em\relax New York, NY, USA: Association for Computing Machinery, Jul. 2019, pp. 1--2.

\bibitem{Ryu2020-el}
N.~Ryu, W.~Lee, M.~J. Kim, and A.~Bianchi, ``Elastick: A handheld variable stiffness display for rendering dynamic haptic response of flexible object,'' in \emph{Proceedings of the 33rd Annual {ACM} Symposium on User Interface Software and Technology}.\hskip 1em plus 0.5em minus 0.4em\relax New York, NY, USA: Association for Computing Machinery, Oct. 2020, pp. 1035--1045.

\bibitem{Tsai2020-ea}
H.-R. Tsai, C.-W. Hung, T.-C. Wu, and B.-Y. Chen, ``{ElastOscillation}: {3D} multilevel force feedback for damped oscillation on {VR} controllers,'' in \emph{Proceedings of the 2020 {CHI} Conference on Human Factors in Computing Systems}, ser. CHI '20.\hskip 1em plus 0.5em minus 0.4em\relax New York, NY, USA: Association for Computing Machinery, Apr. 2020, pp. 1--12.

\bibitem{Ryu2021-xi}
N.~Ryu, H.-Y. Jo, M.~Pahud, M.~Sinclair, and A.~Bianchi, ``{GamesBond}: Bimanual haptic illusion of physically connected objects for immersive {VR} using grip deformation,'' in \emph{Proceedings of the 2021 {CHI} Conference on Human Factors in Computing Systems}, ser. CHI '21, no. Article 125.\hskip 1em plus 0.5em minus 0.4em\relax New York, NY, USA: Association for Computing Machinery, May 2021, pp. 1--10.

\bibitem{heo2018thor}
S.~Heo, C.~Chung, G.~Lee, and D.~Wigdor, ``Thor's hammer: An ungrounded force feedback device utilizing {Propeller-Induced} propulsive force,'' in \emph{Proceedings of the 2018 {CHI} Conference on Human Factors in Computing Systems}, ser. CHI '18, no. Paper 525.\hskip 1em plus 0.5em minus 0.4em\relax New York, NY, USA: Association for Computing Machinery, Apr. 2018, pp. 1--11.

\bibitem{10.1145/3332165.3347926}
S.~Je, M.~J. Kim, W.~Lee, B.~Lee, X.-D. Yang, P.~Lopes, and A.~Bianchi, ``Aero-plane: A handheld {Force-Feedback} device that renders weight motion illusion on a virtual {2D} plane,'' in \emph{Proceedings of the 32nd Annual {ACM} Symposium on User Interface Software and Technology}, ser. UIST '19.\hskip 1em plus 0.5em minus 0.4em\relax New York, NY, USA: Association for Computing Machinery, Oct. 2019, pp. 763--775.

\bibitem{Sasaki:2018:LMH:3214907.3214913}
\BIBentryALTinterwordspacing
T.~Sasaki, R.~S. Hartanto, K.-H. Liu, K.~Tsuchiya, A.~Hiyama, and M.~Inami, ``{Leviopole: Mid-air Haptic Interactions Using Multirotor},'' in \emph{ACM SIGGRAPH 2018 Emerging Technologies}, ser. SIGGRAPH '18.\hskip 1em plus 0.5em minus 0.4em\relax New York, NY, USA: ACM, 2018, pp. 12:1----12:2. [Online]. Available: \url{http://doi.acm.org/10.1145/3214907.3214913}
\BIBentrySTDinterwordspacing

\bibitem{Ke2023-qk}
P.~Ke, S.~Cai, H.~Gao, and K.~Zhu, ``\BIBforeignlanguage{en}{{PropelWalker}: A {Leg-Based} wearable system with {Propeller-Based} force feedback for walking in fluids in {VR}},'' \emph{\BIBforeignlanguage{en}{IEEE Trans. Vis. Comput. Graph.}}, vol.~29, no.~12, pp. 5149--5164, Dec. 2023.

\bibitem{Wang2021-sd}
Y.-W. Wang, Y.-H. Lin, P.-S. Ku, Y.~Miyatake, Y.-H. Mao, P.~Y. Chen, C.-M. Tseng, and M.~Y. Chen, ``{JetController}: High-speed ungrounded {3-DoF} force feedback controllers using air propulsion jets,'' in \emph{Proceedings of the 2021 {CHI} Conference on Human Factors in Computing Systems}, ser. CHI '21, no. Article 124.\hskip 1em plus 0.5em minus 0.4em\relax New York, NY, USA: Association for Computing Machinery, May 2021, pp. 1--12.

\bibitem{Tsai2022-sz}
C.-Y. Tsai, I.-L. Tsai, C.-J. Lai, D.~Chow, L.~Wei, L.-P. Cheng, and M.~Y. Chen, ``{AirRacket}: Perceptual design of ungrounded, directional force feedback to improve virtual racket sports experiences,'' in \emph{{CHI} Conference on Human Factors in Computing Systems}, ser. CHI '22, no. Article 185.\hskip 1em plus 0.5em minus 0.4em\relax New York, NY, USA: Association for Computing Machinery, Apr. 2022, pp. 1--15.

\bibitem{Winfree2009-rv}
K.~N. Winfree, J.~Gewirtz, T.~Mather, J.~Fiene, and {others}, ``A high fidelity ungrounded torque feedback device: The {iTorqU} 2.0,'' \emph{World Haptics 2009}, 2009.

\bibitem{Walker_2018}
J.~M. Walker, H.~Culbertson, M.~Raitor, and A.~M. Okamura, ``\BIBforeignlanguage{en}{Haptic orientation guidance using two parallel {Double-Gimbal} control moment gyroscopes},'' \emph{\BIBforeignlanguage{en}{IEEE Trans. Haptics}}, vol.~11, no.~2, pp. 267--278, Apr. 2018.

\bibitem{nakamura2005innovative}
N.~Nakamura and Y.~Fukui, ``{An innovative non-grounding haptic interface'GyroCubeSensuous' displaying illusion sensation of push, pull and lift},'' in \emph{ACM SIGGRAPH 2005 Posters}.\hskip 1em plus 0.5em minus 0.4em\relax ACM, 2005, p.~92.

\bibitem{1191223}
H.~Yano, M.~Yoshie, and H.~Iwata, ``{Development of a non-grounded haptic interface using the gyro effect},'' in \emph{11th Symposium on Haptic Interfaces for Virtual Environment and Teleoperator Systems, 2003. HAPTICS 2003. Proceedings.}, mar 2003, pp. 32--39.

\bibitem{Hashimoto2022-af}
T.~Hashimoto, S.~Yoshida, and T.~Narumi, ``{MetamorphX}: An ungrounded {3-DoF} moment display that changes its physical properties through rotational impedance control,'' in \emph{Proceedings of the 35th Annual {ACM} Symposium on User Interface Software and Technology}, ser. UIST '22, no. Article 72.\hskip 1em plus 0.5em minus 0.4em\relax New York, NY, USA: Association for Computing Machinery, Oct. 2022, pp. 1--14.

\bibitem{Hashimoto2023-ah}
{T. Hashimoto}, S.~Yoshida, and T.~Narumi, ``{SomatoShift}: A wearable haptic display for somatomotor reconfiguration via modifying acceleration of body movement,'' in \emph{{ACM} {SIGGRAPH} 2023 Emerging Technologies}, ser. SIGGRAPH '23, no. Article 17.\hskip 1em plus 0.5em minus 0.4em\relax New York, NY, USA: Association for Computing Machinery, Jul. 2023, pp. 1--2.

\bibitem{Lopes2017}
P.~Lopes, S.~You, L.-P. Cheng, S.~Marwecki, and P.~Baudisch, ``{Providing Haptics to Walls {\&} Heavy Objects in Virtual Reality by Means of Electrical Muscle Stimulation},'' pp. 1471--1482, 2017.

\bibitem{Lopes2015}
P.~Lopes, A.~Ion, and P.~Baudisch, ``Impacto: Simulating p hysical i mpact by combining tactile stimulation with electrical mu scle stimulation,'' \emph{UIST; Proceedings of the Annual ACM Symposium on User Interface Software and Technology}, pp. 11--19.

\bibitem{Shimizu2021-df}
S.~Shimizu, T.~Hashimoto, S.~Yoshida, R.~Matsumura, T.~Narumi, and H.~Kuzuoka, ``Unident: Providing impact sensations on handheld objects via {High-Speed} change of the rotational inertia,'' in \emph{2021 {IEEE} Virtual Reality and {3D} User Interfaces ({VR})}, Mar. 2021, pp. 11--20.

\bibitem{4600289}
T.~Amemiya and T.~Maeda, ``Asymmetric oscillation distorts the perceived heaviness of handheld objects,'' \emph{IEEE Transactions on Haptics}, vol.~1, no.~1, pp. 9--18, 2008.

\bibitem{hogan1985impedance}
N.~Hogan, ``\BIBforeignlanguage{en}{Impedance control: An approach to manipulation: Part {II---Implementation}},'' \emph{\BIBforeignlanguage{en}{J. Dyn. Syst. Meas. Control}}, vol. 107, no.~1, pp. 8--16, Mar. 1985.

\bibitem{braun2006using}
V.~Braun and V.~Clarke, ``Using thematic analysis in psychology,'' \emph{Qualitative research in psychology}, vol.~3, no.~2, pp. 77--101, 2006.

\bibitem{Shirado2004KJ00007553977}
H.~Shirado and T.~Maeno, ``Modeling of texture perception mechanism for tactile display and sensor,'' \emph{Transactions of the Virtual Reality Society of Japan}, vol.~9, no.~3, pp. 235--240, 2004.

\end{thebibliography}

\section{Biography Section}
 
\vspace{-11pt}

\begin{IEEEbiography}[{\includegraphics[width=1in,height=1.25in,clip,keepaspectratio]{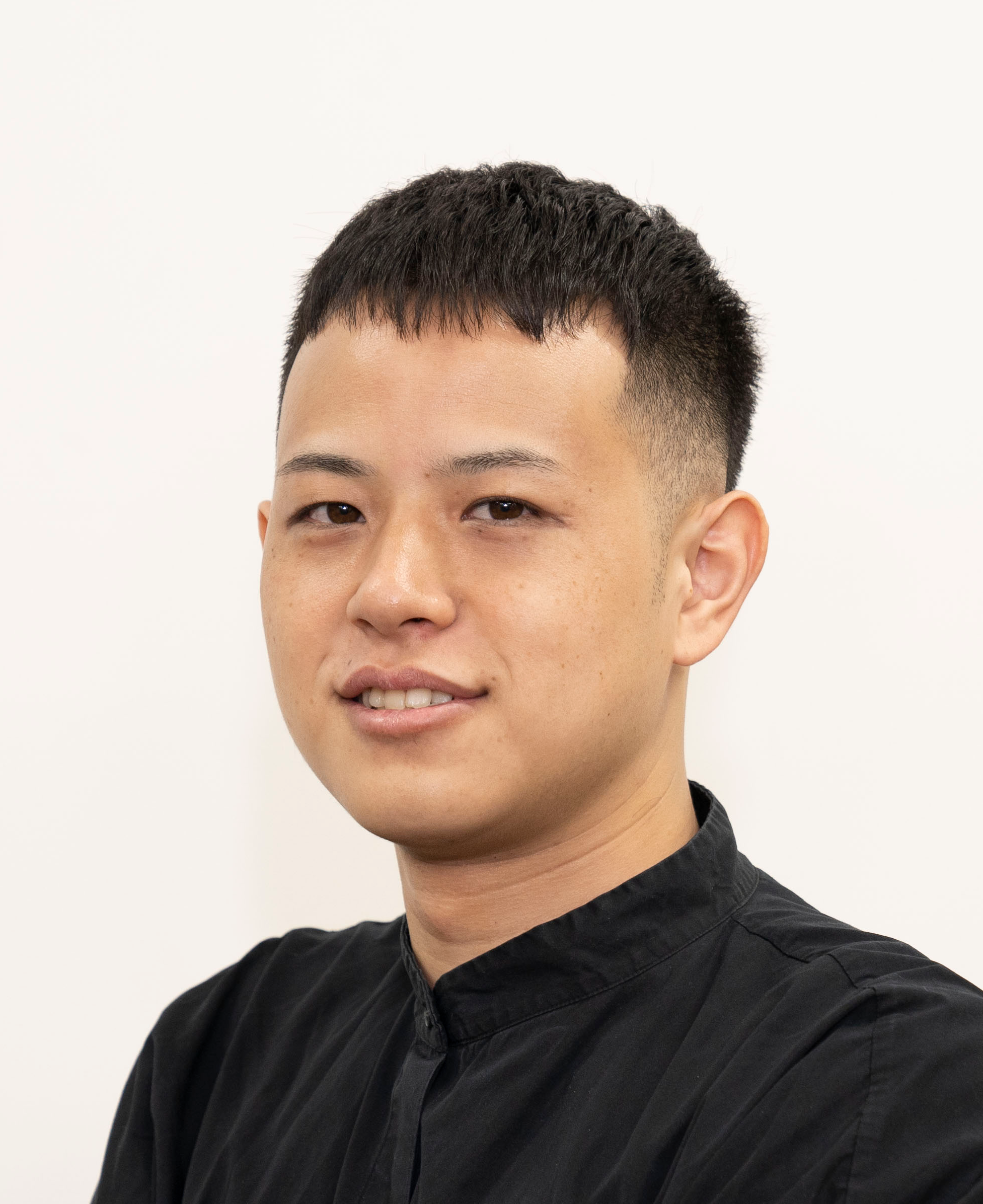}}]{Takeru Hashimoto} is currently a project assistant professor at the Cyber Interface Laboratory in The University of Tokyo in Japan. He received his Ph.D. in information science and technology from The University of Tokyo in 2023. His awards include the Young Researcher’s Award from the virtual reality society of Japan and the Honorable Mentions Award in SIGCHI. His current research area is Virtual / Augmented Reality of haptics and Human-Augmentation.

\end{IEEEbiography}
\vspace{-11pt}

\begin{IEEEbiography}[{\includegraphics[width=1in,height=1.25in,clip,keepaspectratio]{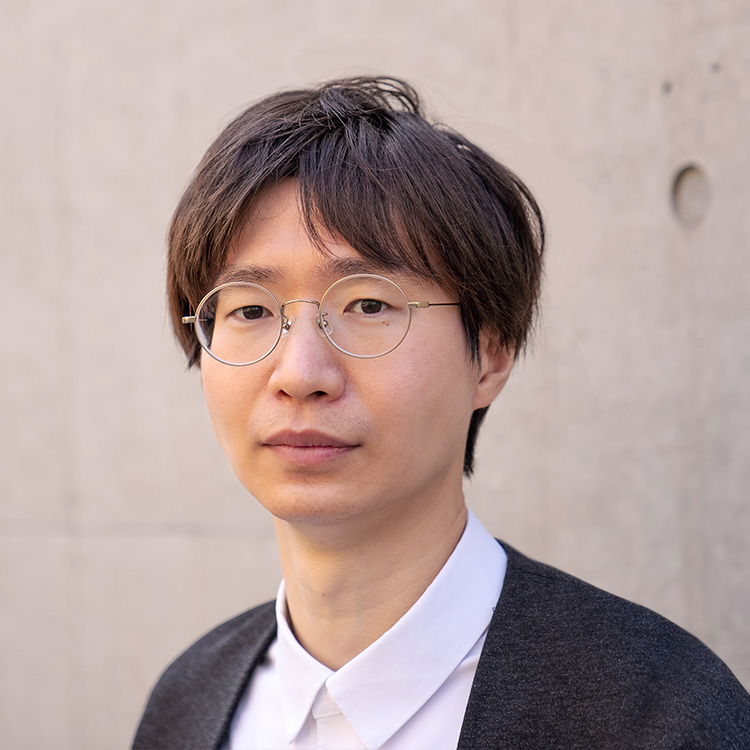}}]{Shigeo Yoshida} is currently a senior researcher at OMRON SINIC X Corporation in Japan. He received his Ph.D. in Information Studies, master's degree in Arts and Sciences, and bachelor's degree in Engineering from The University of Tokyo in 2017, 2014 and 2012 respectively.
His research interest involves a broad area of Human-Computer Interaction. He has been especially focusing on designing interactions based on the mechanisms of perception and cognition of our body.
\end{IEEEbiography}
\vspace{-11pt}

\begin{IEEEbiography}[{\includegraphics[width=1in,height=1.25in,clip,keepaspectratio]{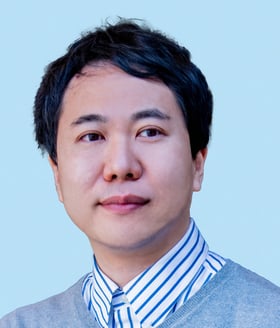}}]{Takuji Narumi} received the B.E. and M.E. degrees from The University of Tokyo, in 2006 and 2008, respectively, and the Ph.D. degree in engineering from The University of Tokyo, in 2011. He is currently an Associate Professor with the Graduate School of Information Science and Technology, The University of Tokyo.
His research interests include the intersection of technologies and human science, and he has been working on extending human senses, cognition, and behavior by combining virtual reality and augmented reality technologies with findings from psychology and cognitive science. He was a recipient of several awards, including MEXT The Young Scientists’ Award, the SIGCHI Japan Chapter Distinguished Young Researcher Award, the Japan Media Arts Festival 2017 Excellence Award, and the CHI Honorable
Mentions, in 2019 and 2020.
\end{IEEEbiography}

\end{document}